\documentclass[11pt,twoside,nohyper]{pallis}
\usepackage{epsfig}
\usepackage{latexsym}
\usepackage{amssymb}
\usepackage{times}
\usepackage{caption}
\usepackage{slashed,cite}
\usepackage{upgreek}

\usepackage{amsmath}
\usepackage{amsfonts}
\usepackage{amsbsy}
\usepackage{amscd}
\usepackage{bbm}

\newcommand{\bal}{\begin{align}}
\newcommand{\eal}{\end{align}}
\newcommand{\beqs}{\begin{subequations}}
\newcommand{\eeqs}{\end{subequations}}
\newcommand{\ec}{\end{center}}
\newcommand{\bec}{\begin{center}}

\newcommand{\eem}{\end{matrix}}
\newcommand{\bem}{\begin{matrix}}
\newcommand{\eeq}{\end{equation}}
\newcommand{\beq}{\begin{equation}}
\newcommand{\ba}{\begin{array}}
\newcommand{\ea}{\end{array}}
\newcommand{\bea}{\begin{eqnarray}}
\newcommand{\eea}{\end{eqnarray}}
\newcommand{\baq}{\begin{eqnarray}}
\newcommand{\eaq}{\end{eqnarray}}

\newcommand{\Eref}[1]{Eq.~(\ref{#1})}
\newcommand{\Sref}[1]{Sec.~\ref{#1}}
\newcommand{\Fref}[1]{Fig.~\ref{#1}}
\newcommand{\Tref}[1]{Table~\ref{#1}}
\newcommand{\cref}[1]{Ref.~\cite{#1}}

\newcommand\eqs[2]{Eqs.~(\ref{#1}) and (\ref{#2})}

\newcommand\eqss[3]{Eqs.~(\ref{#1}), (\ref{#2}) and (\ref{#3})}

\newcommand{\ftn}{\footnotesize}

\newcommand{\ssz}{\scriptsize}

\newcommand{\GeV}{{\mbox{\rm GeV}}}

\newcommand{\sFref}[2]{Fig.~\ref{#1}-{\ftn\sf ({#2})}}
\newcommand{\sEref}[2]{Eq.~(\ref{#1}{\ftn\sf {#2}})}

\newcommand{\etal}{{\it et al.\/}}

\def\to{\rightarrow}

\def\lf{\left(}
\def\rg{\right)}
\newcommand\vev[1]{\langle {#1} \rangle}

\newcommand{\Vhi}{\ensuremath{\widehat V_{\rm CI}}}
\newcommand{\dV}{\ensuremath{\Delta\widehat V_{\rm CI}}}
\newcommand{\Hhi}{\ensuremath{\widehat H_{\rm CI}}}
\newcommand{\Ohi}{\ensuremath{\Omega}}
\newcommand{\Omg}{\ensuremath{\Omega}}
\newcommand{\Khi}{\ensuremath{K}}
\newcommand{\Whi}{\ensuremath{W}}
\newcommand{\Vhio}{\ensuremath{\widehat V_{\rm CI0}}}

\newcommand{\mP}{\ensuremath{M_{\rm P}}}

\newcommand{\Qef}{\ensuremath{\Lambda_{\rm UV}}}

\def\openone{\leavevmode\hbox{\small1\kern-3.8pt\normalsize1}}

\newcommand{\kx}{\ensuremath{k_S}}

\newcommand{\Fcr}{\ensuremath{\Omega_{\rm H}}}
\newcommand{\fr}{\ensuremath{f_{\cal R}}}
\newcommand{\fk}{\ensuremath{\Omega_{\rm H}}}
\newcommand{\fkk}{\ensuremath{f_K}}
\newcommand{\re}{\ensuremath{e_n}}

\newcommand{\fsp}{\ensuremath{f_{S\Phi}}}
\newcommand{\Fk}{\ensuremath{\Omega_{\rm K}}}
\newcommand{\ks}{\ensuremath{k_S}}
\newcommand{\ksp}{\ensuremath{k_{S\Phi}}}
\newcommand{\bksp}{\ensuremath{\bar k_{S\Phi}}}
\newcommand{\kpp}{\ensuremath{k_{\Phi}}}
\newcommand{\gx}{\ensuremath{g_{\chi}}}
\newcommand{\gy}{\ensuremath{g_{\psi}}}

\newcommand{\ck}{\ensuremath{c_{\cal R}}}

\newcommand{\msn}{\ensuremath{\what m_{\rm \dph}}}

\newcommand{\ns}{\ensuremath{n_{\rm s}}}
\newcommand{\as}{\ensuremath{a_{\rm s}}}
\newcommand{\As}{\ensuremath{A_{\rm s}}}
\newcommand{\rcc}{\ensuremath{\mathcal{R}}}
\newcommand{\rce}{\ensuremath{\widehat{\mathcal{R}}}}
\newcommand{\Ve}{\ensuremath{\widehat{V}}}
\newcommand{\He}{\ensuremath{{\what H}}}
\newcommand{\Ne}{\ensuremath{{\what N}}}
\newcommand{\Ns}{\ensuremath{\what N_{\star}}}

\newcommand{\dphi}{\ensuremath{\what{\delta\phi}}}
\newcommand{\dph}{\ensuremath{\delta\phi}}

\newcommand{\what}{\ensuremath{\widehat}}

\newcommand{\Lg}{\ensuremath{\mathcal{L}}}
\newcommand{\Lge}{\ensuremath{\widehat{\mathcal{L}}}}

\def\aal{{\bar\alpha}}
\def\bbet{{\bar\beta}}
\def\al{{\alpha}}

\def\bt{{\beta}}

\def\th{{\theta}}

\newcommand{\Trh}{\ensuremath{T_{\rm rh}}}
\newcommand{\sg}{\ensuremath{\phi}}

\newcommand{\sgx}{\ensuremath{\phi_\star}}
\newcommand{\sgf}{\ensuremath{\phi_{\rm f}}}
\newcommand{\xsg}{\ensuremath{\phi}}

\newcommand{\Ld}{\ensuremath{\Lambda}}
\newcommand{\kp}{\ensuremath{\kappa}}
\newcommand{\se}{\ensuremath{\widehat \phi}}
\newcommand{\sex}{\ensuremath{\widehat{\phi}_\star}}

\newcommand{\geu}{\ensuremath{\widehat g}}
\newcommand{\eph}{\ensuremath{\widehat \epsilon}}
\newcommand{\ith}{\ensuremath{\widehat \eta}}

\newcommand\Gm[1]{\widehat\Gamma_{#1}}

\def\trns{transplanckian}
\def\Ka{K\"{a}hler potential}
\def\Km{K\"{a}hler manifold}

\def\sub{subplanckian}

\newcommand{\kns}{\ensuremath{k_{\sf NS}}}
\renewcommand{\arg}{\ensuremath{{\small\sf arg}}}

\newcommand{\bicep}{{B{\scshape icep}2}}
\newcommand{\plk}{\emph{Planck}}

\renewenvironment{subequations}{%
\refstepcounter{equation}%
\setcounter{parentequation}{\value{equation}}%
  \setcounter{equation}{0}
  \def\theequation{\thesection.\theparentequation{\sf\ftn \alph{equation}}}%
  \ignorespaces
}{%
  \setcounter{equation}{\value{parentequation}}%
  \ignorespacesafterend
}

\title{\Large \bfseries
GRAVITY WAVES FROM NON-MINIMAL QUADRATIC INFLATION}

\author{\large \bfseries\scshape Constantinos Pallis$^{\sf(1)}$ and Qaisar Shafi$^{\sf(2)}$\\

$^{\sf(1)}$ Departament de F\'isica Te\`orica and IFIC,
Universitat de Val\`encia-CSIC, \\ E-46100 Burjassot, SPAIN; {\sl
e-mail address: }{\ftn\tt cpallis@ific.uv.es}\\\vspace{3pt}
$^{\sf(2)}$ Bartol Research Institute, Department of Physics and
Astronomy, University of Delaware, \\ Newark, DE 19716, USA; {\sl
e-mail address: }{\ftn\tt shafi@bartol.udel.edu}}

\abstract{We discuss non-minimal quadratic inflation in
supersymmetric (SUSY) and non-SUSY models which entails a linear
coupling of the inflaton to gravity. Imposing a lower bound on the
parameter $\ck$, involved in the coupling between the inflaton and
the Ricci scalar curvature, inflation can be attained even for
\sub\ values of the inflaton while the corresponding effective
theory respects the perturbative unitarity up to the Planck scale.
Working in the non-SUSY context we also consider radiative
corrections to the inflationary potential due to a possible
coupling of the inflaton to bosons or fermions. We find ranges of
the parameters, depending mildly on the renormalization scale,
with adjustable values of the spectral index $\ns$,
tensor-to-scalar ratio $r\simeq(2-4)\cdot10^{-3}$, and an inflaton
mass close to $3\cdot10^{13}~\GeV$. In the SUSY framework we
employ two gauge singlet chiral superfields, a logarithmic \Ka\
including all the allowed terms up to fourth order in powers of
the various fields, and determine uniquely the superpotential by
applying a continuous $R$ and a global $U(1)$ symmetry. When the
\Km\ exhibits a no-scale-type symmetry, the model predicts
$\ns\simeq0.963$ and $r\simeq0.004$. Beyond no-scale SUGRA, $\ns$
and $r$ depend crucially on the coefficient involved in the fourth
order term, which mixes the inflaton with the accompanying
non-inflaton field in the \Ka, and the prefactor  encountered in
it. Increasing slightly the latter above $(-3)$, an efficient
enhancement of the resulting $r$ can be achieved putting it in the
observable range. The inflaton mass in the last case is confined
in the range $(5-9)\cdot10^{13}~\GeV$.
\\ \\
{\ftn \sf Keywords: Cosmology, Supersymmetric models, Supergravity};\\
{\ftn \sf PACS codes: 98.80.Cq, 11.30.Qc, 12.60.Jv, 04.65.+e}
\\\\{\sl\bfseries Published in} {\sl J. Cosmol. Astropart. Phys. \textbf{03},
023 (2015)}}

\begin{document}

\addtolength{\headheight}{.5cm}

%\tableofcontents\vskip-1.3cm\noindent\rule\textwidth{.4pt}%\\
%\vspace*{.3cm}

\section{Introduction}\label{intro} %

The simplest model \cite{chaotic} of \emph{chaotic inflation}
({\ftn\sf CI}) based on a quadratic potential predicts a (scalar)
spectral index $\ns\simeq0.963$ (in good agreement with WMAP
\cite{wmap} and \plk\ \cite{plin}  measurements) and a
tensor-to-scalar ratio $r$, a canonical measure of primordial
gravity waves, close to $0.15$ or so. The \bicep\ results
\cite{gws} announced earlier this year, purporting to have found
gravity waves from inflation ($r\simeq0.16$) provided a huge boost
for this class of models \cite{oss1, oss, shift1, rStar, rEllis}.
However, serious doubts regarding the \bicep\ results have
appeared in the literature \cite{gws1,gws2} that are largely
related to the inadequate treatment of the impact on their
analysis of the dust background. Furthermore, very recently, the
\plk\ HFI 353 GHz dust polarization data \cite{pdust} has been
released and the first attempts to make a joint analysis of \plk\
and \bicep\ data have been presented \cite{gws2, kdust} concluding
that the quadratic CI is disfavored at more than 95$\%$
\emph{confidence level} ({\sf\ftn c.l.}). Indeed, it is
conceivable that most, if not the whole, \bicep\  polarization
signal may be caused by the dust.

Be that as it may, it was shown several years ago \cite{circ} that
a quadratic (or quartic) potential can, at best, function as an
approximation within a more realistic inflationary cosmology. The
end of CI is followed by a reheating phase which is implemented
through couplings involving the inflaton and some additional
suitably selected fields. The presence of these additional
couplings can significantly modify, through \emph{radiative
corrections} ({\ftn\sf RCs}), the tree level inflationary
potential. For instance, for a quadratic potential supplemented by
a coupling of the inflation field to, say, right-handed neutrinos,
$r$ can be reduced to values close to $0.05$ \cite{oss1} or so, at
the cost of a (less efficient) reduction of $\ns$, though. In this
paper we briefly review this idea taking into account the recent
refinements of \cref{Qenq}, according to which an unavoidable
dependence of the results on the renormalization scale arises.

Another mechanism for reducing $r$ at an acceptable level within
models of quadratic CI is the introduction of a strong, linear
non-minimal coupling of the inflaton to gravity \cite{nmi,nMCI}.
The aforementioned mechanism, that we mainly pursue here, can be
applied either within a \emph{supersymmetric} ({\ftn\sf SUSY})
\cite{nMCI} or a non-SUSY \cite{nmi} framework. The resulting
inflationary scenario, named \emph{non-minimal CI} ({\sf\ftn
nMI}), belongs to a class of universal ``attractor'' models
\cite{roest}, in which an appropriate choice of the non-minimal
coupling to gravity suitably flattens the inflationary potential,
such that $r$ is heavily reduced but $\ns$ stays close to the
currently preferred value of $0.96$. However, in generic
\emph{Supergravity} ({\ftn\sf SUGRA}) settings, a mild tuning is
needed \cite{talk} respecting the coefficient $\ksp$ involved in
the fourth order term that mixes the inflaton with the
accompanying non-inflaton field in the \Ka.

In this work we reexamine the realization of nMI based on the
quadratic potential implementing the following improvements:

\begin{itemize}

\item As regards the non-SUSY case, we also consider RCs to the
tree-level potential which arise due to Yukawa interactions of the
inflaton -- cf.~\cref{shap, ld}. We show that the presence of RCs
can affect the $\ns$ values of nMI -- in contrast to minimal CI,
where RCs influence both $\ns$ and $r$. For \sub\ values of the
inflaton field, though, $r$ remains well suppressed and may be
observable only in the next generation of experiments such as
COrE+ \cite{core}, PIXIE \cite{pixie} and LiteBIRD \cite{lite}
which may bring the sensitivity down to $10^{-3}$.

\item As regards the SUSY case, following \cref{nIG}, we
generalize the embedding of the model in SUGRA allowing for a
variation of the numerical prefactor encountered in the adopted
\Ka. We show that (i) the tuning of $\ksp$ can be totally avoided
in the case of no-scale SUGRA which uniquely predicts
$\ns\simeq0.963$ and $r\simeq0.004$; (ii) beyond no-scale SUGRA,
increasing slightly the prefactor $(-3)$ encountered in the
adopted \Ka\ and adjusting appropriately $\ksp$, an efficient
enhancement of the resulting $r$, for any $\ns$, can be achieved
which will be tasted in the near future \cite{plnext, cmbpol}.

\end{itemize}
We finally show that, in both of the above cases, the
\emph{ultaviolet} ({\ftn\sf UV}) cut-off scale \cite{cutoff,cutof}
of the theory can be identified with the Planck scale and, thus,
concerns regarding the naturalness of this kind of nMI can be
safely evaded. It is worth emphasizing that this nice feature of
these models was recently noticed in \cref{riotto} and was not
recognized in the original papers \cite{nmi,nMCI}.

The paper is organized as follows: In Sec.~\ref{gen}, we describe
the generic formulation of CI with a quadratic potential and a
non-minimal coupling to gravity. The emergent non-SUSY and SUSY
inflationary models are analyzed in Secs.~\ref{nmi} and \ref{fhi}
respectively. The UV behavior of these models is analyzed in
Sec.~\ref{fhi3} and our conclusions are summarized in
Sec.~\ref{con}. In Appendix A we outline the implementation of nMI
by the imaginary part of the inflaton superfield adopting a
shift-symmetric logarithmic \Ka. Throughout the text, the symbol
$,\chi$ as subscript denotes derivation \emph{with respect to}
({\ftn\sf w.r.t}) the field $\chi$ (e.g.,
$_{,\chi\chi}=\partial^2/\partial\chi^2$); charge conjugation is
denoted by a star, and we use units where the reduced Planck scale
$\mP = 2.43\cdot 10^{18}~\GeV$ is set equal to unity.

\section{Inflaton non-Minimally Coupled to Gravity}\label{gen}

We consider below an inflationay sector coupled non-minimally to
gravity within a non-SUSY (\Sref{gen1}) or a SUSY (\Sref{gen2})
framework. Based on this formulation, we then derive the
inflationary observables and impose the relevant observational
constraints in \Sref{obs}.

\subsection{non-SUSY Framework}\label{gen1}

In the \emph{Jordan frame} ({\ftn\sf JF}) the action of an
inflaton $\sg$ with potential $V_{\rm CI} \left(\phi\right)$
non-minimally coupled to the Ricci scalar $\rcc$ through a
coupling function $\fr(\sg)$ has the form:
\beq \label{action1} {\sf  S} = \int d^4 x \sqrt{-\mathfrak{g}}
\left(-\frac{1}{2} \fr\rcc +\frac{\fkk}{2}g^{\mu\nu}
\partial_\mu \sg\partial_\nu \sg- V_{\rm CI0} +\Lg_{\rm int}\right),
\eeq
where $\mathfrak{g}$ is the determinant of the background
Friedmann-Robertson-Walker metric, $g^{\mu\nu}$. We allow also for
a kinetic mixing through the function $\fkk(\phi)$ and a part of
the langrangian $\Lg_{\rm int}$ which is responsible for the
interaction of $\phi$ with a boson $\chi$ and a fermion $\psi$,
i.e.,
\beq\Lg_{\rm
int}=\frac12\gx\phi^2\chi^2+\gy\phi\bar\psi\psi\,.\label{lint}\eeq

By performing a conformal transformation \cite{nmi} according to
which we define the \emph{Einstein frame} ({\sf\ftn EF})  metric
\beq\label{contr}
\geu_{\mu\nu}=\fr\,g_{\mu\nu}~~\Rightarrow~~\left\{\bem
%\begin{array}{rl}
\sqrt{-\what{\mathfrak{g}}}=\fr^2\sqrt{-\mathfrak{g}}\hspace*{0.2cm}\mbox{and}\hspace*{0.2cm}
\geu^{\mu\nu}=g^{\mu\nu}/\fr\, \hfill \cr
\widehat\rcc=\left(\rcc+3\Box\ln \fr+3g^{\mu\nu} \partial_\mu
\fr\partial_\nu \fr/2\fr^2\right)/\fr\, \hfill \cr\eem
%\end{array}
\right.\,,\eeq
where $\Box=\lf
-\mathfrak{g}\rg^{-1/2}\partial_\mu\lf\sqrt{-\mathfrak{g}}\partial^\mu\rg$
and hat is used to denote quantities defined in the EF, we can
write ${\sf S}$ in the EF as follows:
\beq {\sf  S}= \!\int d^4 x
\sqrt{-\what{\mathfrak{g}}}\left(-\frac12
\rce+\frac12\geu^{\mu\nu} \partial_\mu \se\partial_\nu
\se-\Ve_{\rm CI0}+\Lge_{\rm int}\right)\,. \label{action} \eeq
The EF canonically normalized field, $\se$, the EF potential,
$\Ve_{\rm CI}$, and the interaction Langrangian, $\Lge_{\rm int}$
turn out to be:
\beq \label{VJe} \mbox{\ftn\sf
(a)}\>\>\>\left(\frac{d\se}{d\sg}\right)^2=J^2=\frac{\fkk}{\fr}+{3\over2}\left({f_{{R},\sg}\over
\fr}\right)^2,\>\>\>\mbox{\ftn\sf (b)}\>\>\>\Ve_{\rm CI0}=
\frac{V_{\rm CI0}}{\fr^2}~~\mbox{and}~~\mbox{\ftn\sf
(c)}\>\>\>\Lge_{\rm int} = \frac{\Lg_{\rm int}}{\fr^2}\,\cdot\eeq
Taking into account that $\what{\chi}=\fr^{-1/2}\chi$,
$\what{\psi}=\fr^{-3/4}\psi
~~\Rightarrow~~\overline{\what{\psi}}=\fr^{-3/4}{\bar \psi}$
\cite{Maeda}, and that the masses of these particles during CI are
heavy enough such that the dependence of $\fr$ on $\phi$ does not
influence their dynamics, $\Lge_{\rm int}$ can be written as
\beq \Lge_{\rm int}=\frac{\gx\phi^2}{2\fr}\,\what\chi^2+
\frac{\gy\phi}{\sqrt{\fr}}\,\overline{\what{\psi}}\,\what\psi\,.\label{wlint}\eeq

From \Eref{VJe} we infer that convenient choices of $V_{\rm CI}$
and $\fr$ assist us to obtain $\Ve_{\rm CI}$ suitable for
observationally consistent CI. Focusing on quadratic CI and
following \cref{nmi,roest}, we select
\beq\label{Vci} \mbox{\ftn\sf (a)}\>\>\> V_{\rm
CI0}={1\over2}m^2\sg^2\>\>\>\mbox{\ftn\sf (b)}\>\>\>
\fr(\sg)=1+\ck\sg\>\>\>\mbox{and}\>\>\mbox{\ftn\sf
(c)}\>\>\>\fkk=1\eeq
where $m$ is the renormalized mass of the inflaton. For $\ck\gg1$
we observe that a sufficiently flat $\Ve_{\rm CI0}$ through
\sEref{VJe}{b} can be obtained which may decrease $r$ from its
value in (minimal) quadratic CI. On the other hand, the
\emph{vacuum expectation value} ({\ftn\sf v.e.v}) of $\sg$ is
$\vev{\phi}=0$, and the validity of ordinary Einstein gravity is
guaranteed since $\vev{\fr} =1$.

\subsection{SUGRA Framework}\label{gen2}

A convenient implementation of nMI in SUGRA is achieved by
employing two singlet superfields, i.e., $z^\al=\Phi, S$, with
$\Phi$ ($\al=1$) and $S$ ($\al=2)$ being the inflaton and a
``stabilized'' field respectively. The EF action for $z^\al$'s
within SUGRA \cite{linde1} can be written as
\beqs \beq\label{Saction1} {\sf S}=\int d^4x \sqrt{-\what{
\mathfrak{g}}}\lf-\frac{1}{2} \rce +K_{\al\bbet}\geu^{\mu\nu}
\partial_\mu z^\al \partial_\nu z^{*\bbet}-\Ve\rg, \eeq
where the summation is taken over the scalar fields $z^\al$,
$K_{\al\bbet}={K_{,z^\al z^{*\bbet}}}$ with
$K^{\bbet\al}K_{\al\bar \gamma}=\delta^\bbet_{\bar \gamma}$,
$\widehat{\mathfrak{g}}$ is the determinant of the EF metric
$\geu_{\mu\nu}$, $\rce$ is the EF Ricci scalar curvature, $\Ve$ is
the EF F--term SUGRA scalar potential which can be extracted once
the superpotential $\Whi$ and the \Ka\ $\Khi$ have been selected,
by applying the standard formula
\beq \Ve=e^{\Khi}\left(K^{\al\bbet}{\rm F}_\al {\rm
F}^*_\bbet-3{\vert W\vert^2}\right),\>\>\>\mbox{where}\>\>\> {\rm
F}_\al=W_{,z^\al} +K_{,z^\al}W.\label{Vsugra} \eeq \eeqs
Note that D-term contributions to $\Ve$ do not exist since we
consider gauge singlet $z^\al$'s.

A quadratic potential for $\Phi$ in this setting can be realized
if we adopt the following superpotential
\beq \label{Whi} \Whi =m S\Phi\,.\eeq
To protect the form of $\Whi$ from higher order terms we impose
two symmetries: Firstly, an $R$ symmetry under which $S$ and
$\Phi$ have charges $1$ and $0$ respectively, which ensures the
linearity of $\Whi$ w.r.t $S$; secondly, a global $U(1)$ symmetry
with assigned charges $-1$ and $1$ for $S$ and $\Phi$
respectively. To verify that $W$ leads to the desired quadratic
potential we present the SUSY limit, $V_{\rm SUSY}$ of $\Ve$,
which is
\beqs\beq V_{\rm SUSY}= m^2\left(|\Phi|^2 +|S|^2\right).
\label{VF}\eeq
Note that the complex scalar components of $\Phi$ and $S$
superfields are denoted by the same symbol. From \Eref{VF}, we can
easily conclude that for $S$ stabilized to zero, $V_{\rm SUSY}$
becomes quadratic w.r.t to the real (or imaginary) part of $\Phi$.
The SUSY vacuum lies at
\beq \vev{S}=\vev{\Phi}=0.\label{vevs} \eeq\eeqs

The construction of \Eref{action1} can be obtained within SUGRA if
we perform the inverse of the conformal transformation described
in \Eref{contr} with
\beq \fr=-\Omega/3(1+n),\label{Omgfr}\eeq
and specify the following relation between $K$ and $\Omega$,
\beq-\Omega/3(1+n)
=e^{-K/3(1+n)}\>\Rightarrow\>K=-3(1+n)\ln\lf-\Omega/3(1+n)\rg\,.\label{Omg1}\eeq
Here $n$ is a dimensionless (small in our approach) parameter
which quantifies the deviation from the standard set-up
\cite{linde1}. Following \cref{nIG} we arrive at the following
action
\beq {\sf S}=\int d^4x
\sqrt{-\mathfrak{g}}\lf\frac{\Omega\rcc}{6(1+n)}+\lf\Omega_{\al{\bbet}}-\frac{n\Omega_{\al}\Omega_{\bbet}}{(1+n)\Omega}\rg\partial_\mu
z^\al \partial^\mu z^{*\bbet}- \frac{\Omega{\cal A}_\mu{\cal
A}^\mu}{(1+n)^3}-V \rg, \label{Sfinal}\eeq
where $V =\Omega^2\Ve/9(1+n)^2$ is the JF potential and ${\cal
A}_\mu$ is \cite{linde1} the purely bosonic part of the on-shell
value of the auxiliary field
\beq {\cal A}_\mu =-i(1+n)\lf \Omega_\al\partial_\mu
z^\al-\Omega_\aal\partial_\mu z^{*\aal}\rg/2\Omega\,.
\label{Acal}\eeq
It is clear from \Eref{Sfinal} that ${\sf S}$ exhibits non-minimal
couplings of the $z^\al$'s to $\rcc$. However, $\Omega$ also
enters the kinetic terms of the $z^\al$'s. To separate the two
contributions we split $\Omega$ into two parts
\beqs\beq -\Omega/3(1+n)=\fk(\Phi)+{\fk}^*(\Phi^*)-\Fk\lf|\Phi|^2,
|S|^2\rg/3(1+n), \label{Omg}\eeq
where $\Fk$ is a dimensionless real function including the kinetic
terms for the $z^\al$'s and takes the form
\beq \label{Fkdef} \Fk\lf|\Phi|^2, |S|^2\rg= {\kns
|\Phi|^2+|S|^2}\,-\, 2\lf\kx|S|^4+\kpp|\Phi|^4+\ksp
|S|^2|\Phi|^2\rg\,,\eeq
with coefficients $\kns, \kx, \kpp$ and $\ksp$ of order unity. The
fourth order term for $S$ is included to cure the problem of a
tachyonic instability occurring along this direction
\cite{linde1}, and the remaining terms of the same order are
considered for consistency -- the factors of $2$ are added just
for convenience. Alternative solutions to the aforementioned
problem of the tachyonic instability are recently identified in
\cref{ketov, nil}. On the other hand, $\Fcr$ in \Eref{Omg} is a
dimensionless holomorphic function which, for $\Fcr>\Fk$,
represents the non-minimal coupling to gravity -- note that
$\Omega_{\al{\bbet}}$ is independent of $\Fcr$ since $\Omg_{{\rm
H},z^\al z^{*\bbet}}=0$. To obtain a situation similar to
\Eref{Vci}, we adopt
\beq \label{Frdef} \Fcr=\frac12+{\ck\over \sqrt{2}}\Phi\,,
\eeq\eeqs
which respects the imposed $R$ symmetry but explicitly breaks
$U(1)$ during nMI. Furthermore, assuming that the phase of $\Phi$,
$\arg\Phi$, is stabilized to zero, the selected $\Fcr$ at the SUSY
vacuum, \Eref{vevs}, reads
\beq -\vev{\Ohi}/3(1+n)=1\,, \label{ig}\eeq
which ensures a recovery of conventional Einstein gravity at the
end of nMI.

When the dynamics of the $z^\al$'s is dominated only by the real
moduli $|z^\al|$, or if $z^\al=0$ for $\al\neq1$ \cite{linde1}, we
can obtain ${\cal A}_\mu=0$ in \Eref{Sfinal}. The choice $n\neq0$,
although not standard, is perfectly consistent with the idea of
nMI. Indeed, the only difference occurring for $n\neq0$ --
compared to the $n=0$ case -- is that the $z^\al$'s do not have
canonical kinetic terms in the JF due to the term proportional to
$\Omg_\al\Omg_\bbet\neq\delta_{\al\bbet}$. This fact does not
cause any problem since the canonical normalization of $\Phi$
keeps its strong dependence on $\ck$ included in $\fk$, whereas
$S$ becomes heavy enough during nMI and so it does not affect the
dynamics -- see \Sref{fhi1}.

In conclusion, through \Eref{Omg1} the resulting  \Ka\ is
\beq  \Khi=-3(1+n)\ln\lf 1+{\ck\over \sqrt{2}}\lf\Phi+\Phi^*\rg-{
|S|^2+\kns|\Phi|^2\over3(1+n)}+2{\kx|S|^4+\kpp|\Phi|^4+\ksp
|S|^2|\Phi|^2\over3(1+n)}\rg.\label{Kol}\eeq
We set $\kns=1$ throughout, except for the case of no-scale SUGRA
which is defined as follows:
\beq
n=0,\>\>\>\kns=0\>\>\>\mbox{and}\>\>\>\ksp=\kpp=0\,.\label{nsks}\eeq
This arrangement, inspired by the early models of soft SUSY
breaking \cite{eno7, noscale}, corresponds to the \Km\ $SU(2,1)/
SU(2)\times U(1)_R$ with constant curvature equal to $-2/3$. In
practice, these choices highly simplify the realization of nMI,
thus rendering it more predictive thanks to a lower number of the
remaining free parameters.

\subsection{Inflationary Observables -- Constraints} \label{obs}

The analysis of nMI can be carried out exclusively in the EF using
the standard slow-roll approximation keeping in mind the
dependence of $\what\phi$ on $\phi$ -- given by \Eref{VJe} in both
the SUSY and non-SUSY set-up. Working this way, in the following
we outline a number of observational requirements with which any
successful inflationary scenario must be compatible -- see, e.g.,
\cref{review}.

\paragraph{\ftn\sf 2.3.1.} The number of e-folds, $\Ns$, that
the scale $k_\star=0.05/{\rm Mpc}$ experiences during CI,
\begin{equation}
\label{Nhi}  \Ns=\int_{\se_{\rm f}}^{\se_\star}\, d\se\:
\frac{\Ve_{\rm CI}}{\Ve_{\rm CI,\se}}= \int_{\sg_{\rm
f}}^{\sg_\star}\, J^2\frac{\Ve_{\rm CI}}{\Ve_{\rm CI,\sg}}{d\sg},
\end{equation}
must be enough to resolve the horizon and flatness problems of
standard big bang, i.e., \cite{liddle, plin}
\begin{equation}  \label{Ntot}
\Ns\simeq61.7+\ln{\what V_{\rm CI}(\sgx)^{1/2}\over\what V_{\rm
CI}(\sgf)^{1/3}}+ {1\over3}\ln T_{\rm
rh}+{1\over2}\ln{\fr(\sgx)\over\fr(\sgf)^{1/3}},
\end{equation}
where $\Ve_{\rm CI}$ is the radiatively corrected EF potential
presented in \Sref{Vnmi} [\Sref{fhi1}] for the non-SUSY [SUSY]
scenario. Also, we assume here that nMI is followed in turn by a
decaying-inflaton, radiation and matter domination, $\Trh$ is the
reheat temperature after nMI, $\sgx~[\se_\star]$ is the value of
$\sg~[\se]$ when $k_\star$ crosses outside the inflationary
horizon, and $\sgf~[\se_{\rm f}]$ is the value of $\sg~[\se]$ at
the end of nMI. The latter can be found, in the slow-roll
approximation for the models considered in this paper, from the
condition
\beqs\beq {\sf
max}\{\widehat\epsilon(\sgf),|\widehat\eta(\sgf)|\}=1,\label{srcon}\eeq
where the slow-roll parameters can be calculated as follows:
\beq \label{sr}\widehat\epsilon= {1\over2}\left(\frac{\Ve_{\rm
CI,\se}}{\Ve_{\rm CI}}\right)^2={1\over2J^2}\left(\frac{\Ve_{\rm
CI,\sg}}{\Ve_{\rm CI}}\right)^2
\>\>\>\mbox{and}\>\>\>\>\>\widehat\eta= \frac{\Ve_{\rm
CI,\se\se}}{\Ve_{\rm CI}}={1\over J^2}\left(\frac{\Ve_{\rm
CI,\sg\sg}}{\Ve_{\rm CI}}-\frac{\Ve_{\rm CI,\sg}}{\Ve_{\rm
CI}}{J_{,\sg}\over J}\right)\cdot \eeq\eeqs
It is worth mentioning that in our approach we calculate $\Ns$
self-consistently with $\Vhi$ and $\Trh$, and do not let it vary
within the interval $50-60$ as often done in the literature -- see
e.g. \cref{plin,oss1}. Our estimation for $\Ns$ in \Eref{Ntot}
takes into account the transition from the JF to EF -- see
\cref{nmi} -- and the assumption that nMI is followed in turn by a
decaying-particle, radiation and matter domination -- for details
see \cref{hinova}. During the first period, we adopt the so-called
\cite{liddle} canonical reheating scenario with an effective
equation-of-state parameter $w_{\rm re}=0$. This value corresponds
precisely to the equation-of-state parameter, $w$, for a quadratic
potential. In the nMI  case we expect that $w$ will deviate
slightly from this value. However, this effect is quite negligible
since for low $\phi$ values the inflationary potential can be well
approximated by a quadratic potential -- see \Sref{fhi3} below.

\paragraph{\ftn\sf 2.3.2.} The amplitude $A_{\rm s}$ of the power spectrum of the curvature perturbation
generated by $\sg$ at the pivot scale $k_\star$ must be consistent
with data~\cite{plin}:
\begin{equation}  \label{Prob}
\sqrt{\As}=\: \frac{1}{2\sqrt{3}\, \pi} \; \frac{\Ve_{\rm
CI}(\sex)^{3/2}}{|\Ve_{\rm
CI,\se}(\sex)|}=\frac{|J(\sg_\star)|}{2\sqrt{3}\, \pi} \;
\frac{\Ve_{\rm CI}(\sg_\star)^{3/2}}{|\Ve_{\rm
CI,\sg}(\sg_\star)|}\simeq4.685\cdot 10^{-5},
\end{equation}
where we assume that no other contributions to the observed
curvature perturbation exists.

\paragraph{\ftn\sf 2.3.3.}  The (scalar) spectral index, $\ns$, its
running, $a_{\rm s}$, and the scalar-to-tensor ratio $r$ must be
in agreement with the fitting of the data \cite{plin} with
$\Lambda$CDM model, i.e.,
\begin{equation}\label{nswmap}
\mbox{\ftn\sf (a)}\>\>\> \ns=0.9603\pm0.0146,\>\>\>\mbox{\ftn\sf
(b)}\>\>\> -0.0314\leq \as\leq0.0046
\>\>\>\mbox{and}\>\>\>\mbox{\ftn\sf (c)}\>\>\>r<0.1\>\>\>\mbox{at
95$\%$.}
\end{equation}
In \sEref{nswmap}{c} we conservatively take into account the
recent analyses \cite{gws2, kdust} which combine the \bicep\
results \cite{gws} with the polarized foreground maps released by
\plk\ \cite{pdust}. These observables are estimated through the
relations:
\beq\label{ns} \mbox{\ftn\sf (a)}\>\>\>\ns=\:
1-6\widehat\epsilon_\star\ +\
2\widehat\eta_\star,\>\>\>\mbox{\ftn\sf (b)}\>\>\>
\as=\:2\left(4\widehat\eta_\star^2-(\ns-1)^2\right)/3-2\widehat\xi_\star\>\>\>
\mbox{and}\>\>\>\mbox{\ftn\sf
(c)}\>\>\>r=16\widehat\epsilon_\star, \eeq
where $\widehat\xi={\Ve_{\rm CI,\widehat\sg} \Ve_{\rm
CI,\widehat\sg\widehat\sg\widehat\sg}/\Ve_{}^2}= \,\Ve_{\rm
CI,\sg}\,\widehat\eta_{,\sg}/\Ve_{\rm
CI}\,J^2+2\widehat\eta\widehat\epsilon$ and the variables with
subscript $\star$ are evaluated at $\sg=\sg_\star$.

\paragraph{\ftn\sf 2.3.4.}  To avoid corrections from quantum
gravity and any destabilization of our inflationary scenario due
to higher order terms -- e.g. in \Eref{Vci} or \Eref{Frdef} --, we
impose two additional theoretical constraints on our models --
keeping in mind that $\Vhi(\sg_{\rm f})\leq\Vhi(\sg_\star)$:
\beq \label{subP}\mbox{\ftn\sf (a)}\>\> \Vhi(\sg_\star)^{1/4}\leq1
\>\>\>\mbox{and}\>\>\>\mbox{\ftn\sf (b)}\>\> \sg_\star\leq1.\eeq
As we show in \Sref{fhi3}, the UV cutoff of our model is $\mP$,
and so concerns regarding the validity of the effective theory are
entirely eliminated.

\section{non-SUSY Inflation}\label{nmi}

Focusing first on the non-SUSY case, we extract the inflationary
potential in \Sref{Vnmi}. Then, to better appreciate the
importance of the non-minimal coupling to gravity for our
scenario, we start the presentation of our results with a brief
revision of the case where the inflaton is minimally coupled to
gravity in \Sref{resnmi1}. We extend our analysis to the more
relevant case of nMI in \Sref{resnmi2}.

\subsection{Inflationary Potential} \label{Vnmi}

The tree-level EF inflationary potential of our model, found by
plugging \Eref{Vci} into \sEref{VJe}{b}, can be supplemented by
the one-loop RCs computed in EF with the use of the standard
formula of \cref{cw} -- cf. \cref{shap}. To this end, we determine
the particle masses as functions of the background field $\sg$ --
see \Eref{wlint}. Our result is
%\begin{multline}
\begin{equation}
\label{dU1loop} \Delta\Vhi = \frac{1}{64\pi^2}\lf{\what
m_{\chi}^4}\ln\frac{\what m_{\chi}^2}{\Ld^2}- 4\what
m_{\psi}^4\ln\frac{\what m_{\psi}^2}{\Ld^2}\rg,\>\>\>
\mbox{with}\>\>\>\what m_{\chi}^2 = \frac{\gx \sg^2}{\fr}\>\>\>
\mbox{and}\>\>\>\what m_{\psi}^2 = \frac{\gy^2 \sg^2}{\fr}\cdot
\eeq
Here $\Ld$ is the renormalization scale and we assume that the
on-shell masses of $\chi$ and $\psi$ are much ligther than the
effective ones. Note that the only difference from the flat space
case \cite{circ,Qenq} is the presence of the conformal factor
$\fr$ in the denominators of the masses. We verify that these
masses are heavier than the Hubble parameter
$\Hhi=(\Vhio/3)^{1/2}$ during CI. On the other hand, the mass of
$\sg$ is much lower than $\Hhi$ and thus, its contribution to
\Eref{dU1loop} can be safely neglected. For numerical
manipulations we find it convenient to write the one-loop
corrected inflationary potential as
\beq\label{Vhirc}
\Vhi=\Vhio+\Delta\Vhi=\frac{m^2\phi^2}{2\fr^2}\lf1+\kp\ln\frac{\sg}{\sqrt{\fr}\Ld}\rg,\>\>\>
\mbox{where}\>\>\>\kp=\frac{\gx^2-4\gy^4}{16\pi^2m^2}
\end{equation}
expresses \cite{Qenq} the inflaton interaction strength. Following
\cref{Qenq} we assume that for $\kp>0$ [$\kp<0$], we have
$\gy\ll\gx$ [$\gx\ll\gy$], and thus $g_\psi$ or $g_\chi$ can be
absorbed by redefining $\Ld$. Since there is no information, from
particle physics about physical quantities -- such as masses and
coupling constants -- which would assist us to determine $\Ld$
uniquely, we consider it as a free parameter and discuss below the
unavoidable dependence of the inflationary predictions on it.

At the end of CI, $\sg$ settles in its v.e.v $\vev{\sg}=0$ and the
EF (canonically normalized) inflaton,
\beq\dphi=\vev{J}\dph\>\>\>\mbox{with}\>\>\>\vev{J}\simeq\sqrt{1+{3\ck^2/2}},
\label{dphi1} \eeq
acquires mass which is given by
\beq \label{msn1} \msn=\left\langle\Ve_{\rm
CI0,\se\se}\right\rangle^{1/2}=m/\vev{J}.\eeq
The decay of $\dphi$ is processed not only through the decay
channel originating from the term in \Eref{wlint} which is
proportional to $\gy$, but also through the spontaneously arisen
interactions which are proportional to $\vev{f_{\cal R,\phi}}=\ck$
\cite{reheating}. The relevant lagrangian which describes these
decay channels reads
\beq \Lge_{\rm dc}=\what g_\psi
\msn\dphi\,\overline{\what{\psi}}\,\what\psi+\,\what
g_\chi\msn\dphi\,\what\chi^2,\>\>\>\mbox{where}\>\>\>\what
g_\psi=\frac{\gy}{\vev{J}}+\frac{\ck
m_\psi}{2\vev{J}}\>\>\>\mbox{and}\>\>\>\what{g}_\chi=\frac{\ck
\msn}{4\vev{J}}\label{Ldecay}\eeq
are dimensionless couplings and $m_\psi$, the mass of $\psi$, is
set equal to $\msn/10$ for numerical applications. As it turns
out, $g_\psi$ dominates the computation of $\what g_\psi$ for all
relevant cases.  These interactions give rise to the following
decay rates of $\dphi$
\beq \Gm{\psi}=\frac{\what{g}_{\psi}^2}{8\pi}\msn
\>\>\>\mbox{and}\>\>\>
\Gm{\chi}=\frac{\what{g}_{\chi}^2}{16\pi}\msn, \label{eq:decay}
\eeq
which can ensure the reheating of the universe with temperature
calculated by the formula \cite{quin}:
\begin{equation}
\Trh\simeq\left(\frac{72}{5\pi^2 g_{*}}\right)^{1/4}
\sqrt{\Gm{\dph}},\>\>\>\mbox{where}\>\>\>\Gm{\dph}=\Gm{\psi}+\Gm{\chi}
\label{GTrh}
\end{equation}
and we set $g_{*}=106.75$ for the relativistic degrees of freedom
assuming the particle spectrum of Standard Model. Summarizing, the
proposed inflationary scenario depends on the parameters:
$$m,\>\ck,\>\kp\>\>\>\mbox{and}\>\>\>\Ld.$$
Following common practice \cite{Qenq}, we consider below two
optimal values which makes $\dV$ vanish for $\sg=\sgx$ or
$\sg=\sgf$.

\subsection{Minimal Coupling to Gravity} \label{resnmi1}

This case can be studied if we set $\fr=1$ and $\fkk=1$, resulting
in $J=1$, in the formulae of Secs.~\ref{gen1}, \ref{Vnmi}, and
\ref{obs} -- hatted and unhatted quantities are identical in this
regime. In our investigation we first extract some analytic
expressions -- see \Sref{mi1} -- which assist us to interpret the
exact numerical results presented in \Sref{mi2}.

\subsubsection{Analytic Results.} \label{mi1} The slow-roll parameters can be calculated by
applying \Eref{sr} with results
\beq\epsilon=\frac12\lf\frac{2 + \kp \sg^2 + 4 \kp
\sg^2\ln\frac{\sg}{\Ld}}{\sg + \kp \sg^3
\ln\frac{\sg}{\Ld}}\rg^2~~~\mbox{and}~~~\eta=\frac{2 + 7 \kp \sg^2
+ 12 \kp \sg^2 \ln\frac{\sg}{\Ld}}{\sg^2 + \kp \sg^4
\ln\frac{\sg}{\Ld}}\cdot\label{misr}\eeq
Numerically we verify that $\sgf$ does not decline by much from
its value for $\kp=0$, i.e., $\sgf\simeq\sqrt{2}$. Hiding the
$\Ld$ dependence, which turns out to be not so significant,
\Eref{Nhi} yields for the number of $e$-foldings experienced from
$k_\star$ during CI
\beq N_\star\simeq \frac1{2 \kp}\ln{\frac{1 + \kp \sgx^2/2}{1 +
\kp}}  ~~\Rightarrow~~~\sgx= \lf\frac{2}{\kp}\lf e^{2 \kp
N_\star}(1 + \kp)-1\rg\rg^{1/2}\cdot\label{miN}\eeq
Note that the above formulae are valid for both signs of $\kp$
although we concentrate below on negative $\kp$ values which
assist us in the reduction of $r$. The normalization of
\Eref{Prob} imposes the condition
\beq  \sqrt{\As} \simeq \frac{m \sgx^2}{2 \sqrt{6} \pi (2 + \kp
\sgx^2)} \>\>\> \Rightarrow\>\>\> m\simeq \frac{2\pi\sqrt{6\As}
e^{2 \kp N_\star} \kp (1 + \kp)}{e^{2\kp N_\star} (1 +
\kp)-1}\cdot\label{miAs}\eeq
In the limit $\kp\to0$, the expressions in \eqs{miN}{miAs} reduce
to the corresponding ones -- see \eqs{im1}{mJ} with $J=1$ -- that
we obtain within the simplest quadratic CI. Upon substitution of
\eqs{misr}{miN} into \Eref{ns} we may compute the inflationary
observables. Namely, \sEref{ns}{a} yields
\beqs\beq\label{mins}\ns\simeq1-\frac{2}{N_\star}+ \left\{\bem
%\begin{array}{rl}
%
4\kp -
38\kp^2N_\star/3+(1/6-12N_\star^2)\kp^3\hfill&\mbox{for}~~\Ld=\sgx\hfill\cr
2(2 - l_\star)\kp  - 4\lf11+ 3 l_\star(7 + 2
l_\star)\rg\kp^2N_\star/3\hfill&\mbox{for}~~\Ld=\sgf\hfill\cr \eem
%\end{array}
\right.\,, \eeq where $l_\star=\ln2N_\star$ and an expansion for
$\kp\ll1$ has been performed. Needless to say, the optimal scale
$\Ld=\sgx$ or $\sgf$ yields $\dV(\sgx)=0$ or $\dV(\sgf)=0$
respectively for $\fr=1$ -- see \Eref{Vhirc}. Similarly, from
\sEref{ns}{b} we get
\beq \as\simeq-\frac{2}{N_\star^2}+ \left\{\bem
%\begin{array}{rl}
%
2\kp/N_\star+ 128\kp^2/3+98 \kp^3 N_\star/3
\hfill&\mbox{for}~~\Ld=\sgx\hfill\cr
2 (1 - 2 l_\star)\kp/N_\star - 4 (29+3l_\star(15+4l_\star))\kp^2
/3 \hfill&\mbox{for}~~\Ld=\sgf\hfill\cr \eem
%\end{array}
\right.\,, \label{mias}\eeq
while \sEref{ns}{c} implies
\beq r\simeq\frac{8}{N_\star}+ \left\{\bem
%\begin{array}{rl}
%
24 \kp + 104 \kp^2 N_\star/3 + 32 \kp^3 N_\star^2
\hfill&\mbox{for}~~\Ld=\sgx\hfill\cr
8 (3 + 4 l_\star)  \kp + 8 (25 - 12l_\star^2)\kp^2N_\star /3
\hfill&\mbox{for}~~\Ld=\sgf\hfill\cr \eem
%\end{array}
\right.\,. \label{mir}\eeq \eeqs
From the expressions above we infer that a negative $\kp$ can
reduce $r$ and, less efficiently, $\ns$ and $|\as|$ below their
values for $\kp=0$.

\subsubsection{Numerical Results.}\label{mi2}  These conclusions are verified numerically in \Tref{tab1} where we
present results compatible with Eqs.~(\ref{Ntot}), (\ref{Prob}),
(\ref{nswmap}{\sf\ftn a, b}) and (\ref{subP}{\sf\ftn a}), taking
$\dV\neq0$ and $\Ld=\sgx$ (cases A and B), or $\Ld=\sgf$ (cases A'
and B') -- note that \sEref{subP}{b} cannot be satisfied. We
observe that by adjusting $|\kp|$ we can succeed to diminish $r$
below its value in quadratic CI without RCs but not a lot lower
than its maximal allowed value in \sEref{nswmap}{c}. Indeed, the
lowest $r$ obtained is $0.054$. Moreover, this reduction causes a
reduction of $\ns$ which acquires its lowest allowed value in
cases A and A' -- see \sEref{nswmap}{a}.  The dependence of the
results on $\Ld$ can be inferred by comparing the sets of
parameters in the primed and unprimed columns. Note that the
reference value of $\ns$ is fixed in every couple of columns --
i.e., $0.946$ in cases A and A' and $0.96$ in cases B and B'. The
$\Ld$-dependence of the results is imprinted mainly on the values
of $\kp$ which are considerably lower for $\Ld=\sgf$. From the
definition of $\gy$ in \Eref{Vhirc}, though, we infer that this
$\Ld$-dependence becomes milder as regards $\gy$ values. Since
$J=1$, we also notice that $\se=\sg$ and $\msn=m$ roughly equal to
its value, $6.8\cdot10^{-6}$, for $\kp=0$.

In conclusion, the consideration of RCs arising from the coupling
of the inflaton to fermions can reconcile somehow $\phi^2$ CI with
data. However, the violation of \sEref{subP}{b} and the
$\Ld$-dependence are two severe shortcomings of this mechanism.

\begin{table}[!t]
\begin{center}
\renewcommand{\arraystretch}{1.3}
{\small \begin{tabular}{|l||ll|ll||ll|ll||l|l|} \hline {\sc Cases}
&A&B&C& D&A' &B' & C'&D'&E&F\\ \hline
\multicolumn{11}{|c|}{\sc Input Parameters}\\\hline\hline
$\sgx$ &$14.1$ &$14.65$&$1$& $0.1$&$13.2$ &$14.2$ & $1$ &$0.1$&$1$&$0.1$\\
$\ck$ &  $0$&$0$&$77$ & $760$&$0$&$0$&$76$ & $730$&$76$&$760$\\
$-\kp/0.01$&$0.34$& $0.1$&$0.17$& $65$&$0.07$& $0.03$&$0.025$&
$10.5$&$0$&$0$
\\ \hline\hline
\multicolumn{11}{|c|}{\sc Output Parameters}\\\hline
&\multicolumn{4}{|c||}{$\dV(\sgx)=0$}&\multicolumn{4}{|c||}{$\dV(\sgf)=0$}
&\multicolumn{2}{|c|}{$\dV=0$}\\\cline{2-11} \hline\hline
$\gy/0.01$ &$0.13$&$0.11$&$1.72$&$23$
&$0.09$&$0.08$&$1.06$&$13.9$&\multicolumn{2}{|c|}{$0$}
\\\cline{10-11}
$m/10^{-3}$ &$0.0048$&$0.006$&$1.15$&$10$ &$0.005$&$0.006$&$1$&$9.5$&$1.2$&$12$ \\
$\sgf/0.1$ &$14.5$ &$14.2$&$0.15$& $0.015$&$14.1$&$14.1$& $0.15$&
$0.015$& $0.15$& $0.015$\\\cline{10-11}
$\Ns$ &  $54.5$&$54.7$&$54.7$& $54.4$&$54.5$&$54.8$& $54.4$&$54.6$&\multicolumn{2}{|c|}{$54.5$}\\
\hline
$\se_\star$ &$14.1$ &$14.65$&$0$& $-2.8$&$13.2$&$14.2$&
$0$&$-2.8$& $0$& $-2.8$\\
$\se_{\rm f}$ &$1.45$ &$1.42$&$-5.1$& $-7.9$&$1.41$&$1.41$&
$-5.1$& $-7.9$& $-5.1$& $-7.9$\\\hline
$\ns/0.1$ &  $9.46$&$9.6$&$9.6$&$9.46$&$9.46$&$9.6$& $9.6$&$9.46$&\multicolumn{2}{|c|}{$9.64$}\\
$-\as/10^{-4}$ &  $2.6$&$6.1$&$3.8$&$-25$&$2.2$&$5.8$& $4.8$&$0.4$&\multicolumn{2}{|c|}{$6.5$}\\
$r/0.1$ &  $0.7$&$1.2$&$0.03$&$0.03$&$0.54$&$1$& $0.03$&$0.025$&\multicolumn{2}{|c|}{$0.04$}\\
\hline
$\msn/10^{-6}$ &  $4.8$&$6$&$12.2$& $11.2$&$5$& $6$&$12.2$&$10.6$&\multicolumn{2}{|c|}{$12.8$}\\
$\Trh/10^{-8}$ &  $19.8$&$18.2$&$4.36$& $5.64$&$14.6$&$13.9$&
$2.7$&$3.5$&\multicolumn{2}{|c|}{$0.065$}\\\hline
\end{tabular}}
\end{center}
\hfill \caption[]{\sl\small Input and output parameters,
compatible with Eqs.~(\ref{Ntot}), (\ref{Prob}),
(\ref{nswmap}{\sffamily \ftn a, b}) and (\ref{subP}{\sffamily \ftn
a}), for minimal (cases A, B, A' and B') and non-minimal (cases C,
D, C' and D') CI and two choices of $\Ld$. For reference, we also
display results for nMI in the absence of RCs (columns
E,F).}\label{tab1}
\end{table}

\subsection{Non-Minimal Coupling to Gravity} \label{resnmi2}

If we employ the linear non-minimal coupling to gravity suggested
in \sEref{Vci}{b} with $\ck\gg1$, we can follow the same steps as
in \Sref{resnmi1} -- see Secs.~\ref{nmi1} and \ref{nmi2} below.

\subsubsection{Analytic Results.} \label{nmi1} From \eqs{VJe}{sr}, we find
\beqs\beq\label{nmisr1} J\simeq\sqrt{\frac{3}{2}}\frac1\sg,~~
\widehat\epsilon=\lf\frac{4 + \kp \sg^2 (2 + \ck \sg) +2 \kp \sg^2
(2 + \ck \sg) \ln\frac{\sg^2}{\Ld^2\fr}}{\sqrt{3}\fr \lf2+\kp
\sg^2\ln\frac{\sg^2}{\Ld^2\fr}\rg}\rg^2,\eeq and \beq
\widehat\eta= 2\frac{8 + \sg (16 \kp \sg + \ck(\kp \sg^2 (15 + 4
\ck \sg)-4)) - 2 \kp \sg^2 (8 +\ck \sg (7 + 2 \ck \sg))
\ln\frac{\sg^2}{\Ld^2\fr}}{3 \fr^2 \lf2+\kp \sg^2
\ln\frac{\sg^2}{\Ld^2\fr}\rg} \cdot\label{nmisr2}\eeq\eeqs
The expressions above reduce to the well known ones
\cite{nmi,talk} for $\kp=0$. We can, also, verify that the
formulas for $\sgf$, $\Ns$ and $\sgx$ found there \cite{nmi,talk}
give rather accurate results even with $\kp\neq0$, i.e.,
\beq\label{nmiN}\Ns\simeq{3\ck\sgx}/{4}~~\Rightarrow~~
{\sgx}\simeq{4\Ns}/{3\ck}\ll{\sgf}\simeq{2}/{\sqrt{3}\ck},\eeq
and $\sgx$ can be subplanckian -- see \sEref{subP}{b} -- if we
confine ourselves to the regime
\beq
\label{res2}\ck\gtrsim4\Ns/3\simeq77\>\>\>\mbox{for}\>\>\>\Ns\simeq54.\eeq
However, $\se$ may be \trns\ since integrating \sEref{VJe}{a} in
view of \Eref{nmisr1} and employing then \Eref{nmiN} we extract
\beq \se=\sqrt{3/2}\ln\sg~\Rightarrow~\left\{\bem
%\begin{array}{rl}
%
\se_\star\simeq \sqrt{3/2}\ln(4\Ns/3\ck)\hfill\cr
\se_{\rm f}\simeq\sqrt{3/2}\ln(2/\sqrt{3}\ck)\hfill\cr \eem
%\end{array}
\right.,\label{se11}\eeq
whose the absolute value is greater than unity for
$\sg\lesssim0.4$. Nonetheless, \sEref{subP}{b} is enough to
protect our scheme from higher order terms. \sEref{subP}{a} does
not restrict the parameters.

The relation between $m$ and $\ck$ implied by \Eref{Prob},
neglecting the $\Ld$ dependence, becomes
\beq \label{Prob2}\sqrt{\As}\simeq \frac{m \sgx}{2\pi (4 + \kp
\sgx^2 (2 + \ck \sgx))}\>\Rightarrow\>m\simeq\frac{2\pi\sqrt{\As}
(27 \ck^2 + 16 \kp \Ns^3)}{9 \ck\Ns}\cdot\eeq
Plugging \eqss{nmisr1}{nmisr2}{nmiN} into \Eref{ns} and expanding
for $\ck\gg1$, we arrive at
\beqs\beq\label{nmins}  \ns\simeq1-\frac{2}{\Ns}+
\frac{128}{27}\frac{\kp\Ns^2}{\ck^2}\delta\ns,\>\>
\as\simeq-\frac{2}{\Ns^2}-
\frac{416}{27}\frac{\kp\Ns}{\ck^2}\delta\as,
\>\>\>\mbox{and}\>\>\>r\simeq\frac{12}{\Ns^2}+
\frac{128}{9}\frac{\kp\Ns}{\ck^2}\delta r,\eeq
where the $\Ld$-dependence is encoded in $\delta\ns,\delta\as$ and
$\delta r$ which are given by
\beq\label{deltaob}  \delta\ns=\left\{\bem
%\begin{array}{rl}
%
1\hfill\cr
1-\what l_\star\hfill\cr \eem
%\end{array}
\right.,~~\delta\as=\left\{\bem
%\begin{array}{rl}
%
1\hfill\cr
1-{10}\what l_\star/{13}\hfill\cr \eem
%\end{array}
\right. ~~\mbox{and}~~\delta r\simeq\left\{\bem
%\begin{array}{rl}
%
1\hfill\cr
1-2\what l_\star\hfill\cr \eem
%\end{array}
\right. ~~\mbox{for} ~~\left\{\bem
%\begin{array}{rl}
%
\dV(\sgx)=0\hfill\cr
\dV(\sgf)=0\hfill\cr \eem
%\end{array}
\right., \eeq\eeqs
where $\what l_\star=-\ln(1+2/\sqrt{3})\Ns$. Since the
$\kp$-dependent correction is proportional to $\Ns^2$ for $\ns$
and just to $\Ns$ for $\as$ and $r$, we expect that $\kp\neq0$ has
a larger impact on $\ns$ and relatively minor on $\as$ and $r$.

\subsubsection{Numerical Results.} \label{nmi2} To emphasize further the salient features of the present model, we
arrange some representative numerical values of its parameters,
fulfilling all the requirements of \Sref{obs}, in columns C, D,
C', D', E  and F of \Tref{tab1}.  More specifically, in columns E
and F we display the predictions of the model if we switch off the
RCs, taking a tiny $\kp$ value. We easily recognize that the
outputs of this model coincide with those of nMI with quartic
potential and quadratic $\fr$ -- see e.g. \cref{plin,shap}.
Therefore, these are in excellent agreement with the current
observational data as regards $\ns$, whereas $r$ is sufficiently
low. If we switch on the RCs and keep $\sgx$ equal to its values
in columns E and F, we note the following: ({\sf\ftn i}) as
anticipated in \Eref{nmins}, adjusting $\kp$ we can reduce $\ns$
whereas the resulting $r$ remains close to its ``universal'' value
in cases E and F; ({\sf\ftn ii}) the resulting $|\as|$ is a little
lower except for case D -- the result is consistent with our
estimate in \Eref{nmins}; ({\sf\ftn iii}) the extracted $m, \sgf,
\se_\star, \se_{\rm f}$ and $\msn$ are close to the corresponding
ones in cases E or F. Comparing the results of columns C, D, C'
and D' with those of A, B, B' and B' we notice that in the
(former) cases with $\ck\neq0$: ({\sf\ftn i}) inflationary
solutions consistent with \sEref{subP}{b} are possible; ({\sf\ftn
ii}) $m$ required by \Eref{Prob} is at most three orders of
magnitude larger whereas \Eref{msn1} yields $\msn$ only one order
of magnitude larger; ({\sf\ftn iii}) the resulting $g_\psi$ is
larger but $\Trh$ is lower; ({\sf\ftn iv}) the resulting $r$ is
almost one order of magnitude smaller; ({\sf\ftn v}) the
$\Ld$-dependence is generally milder. In both cases (minimal and
non-minimal CI), however, the $\kp$ value needed to obtain the
same $\ns$ value is lower for $\dV(\sgf)=0$.

%%%%%%%%%%%%%%%%%%%%%%%%%%%%%%%%%%%%%%%%%%%%%%%%%%%%%%%%%%%%%%%%%%%%%
\begin{figure}[!t]\vspace*{-.12in}
\hspace*{-.15in}
\begin{minipage}{8in}
\epsfig{file=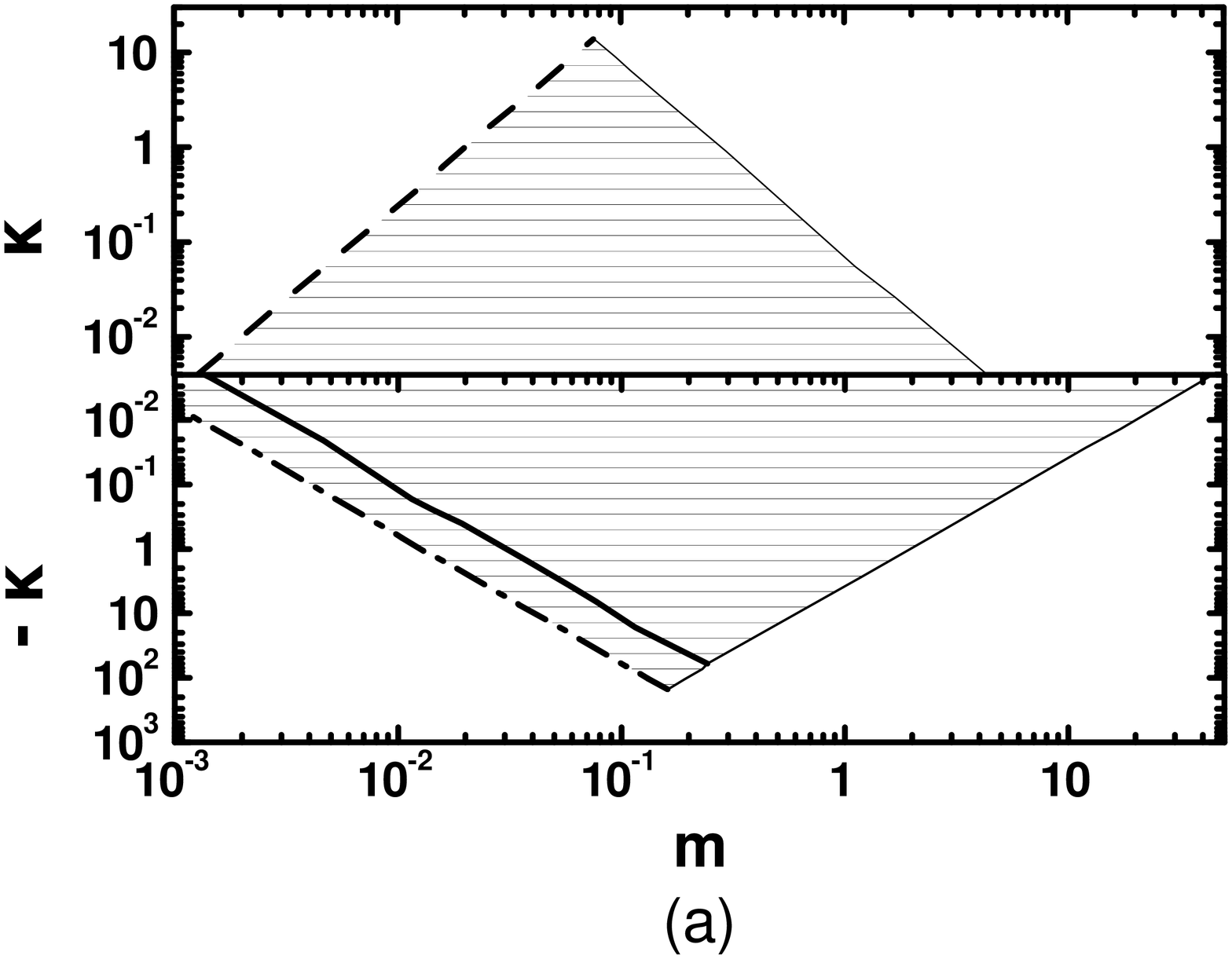,height=3.6in,angle=-90}
\hspace*{-1.2cm}
\epsfig{file=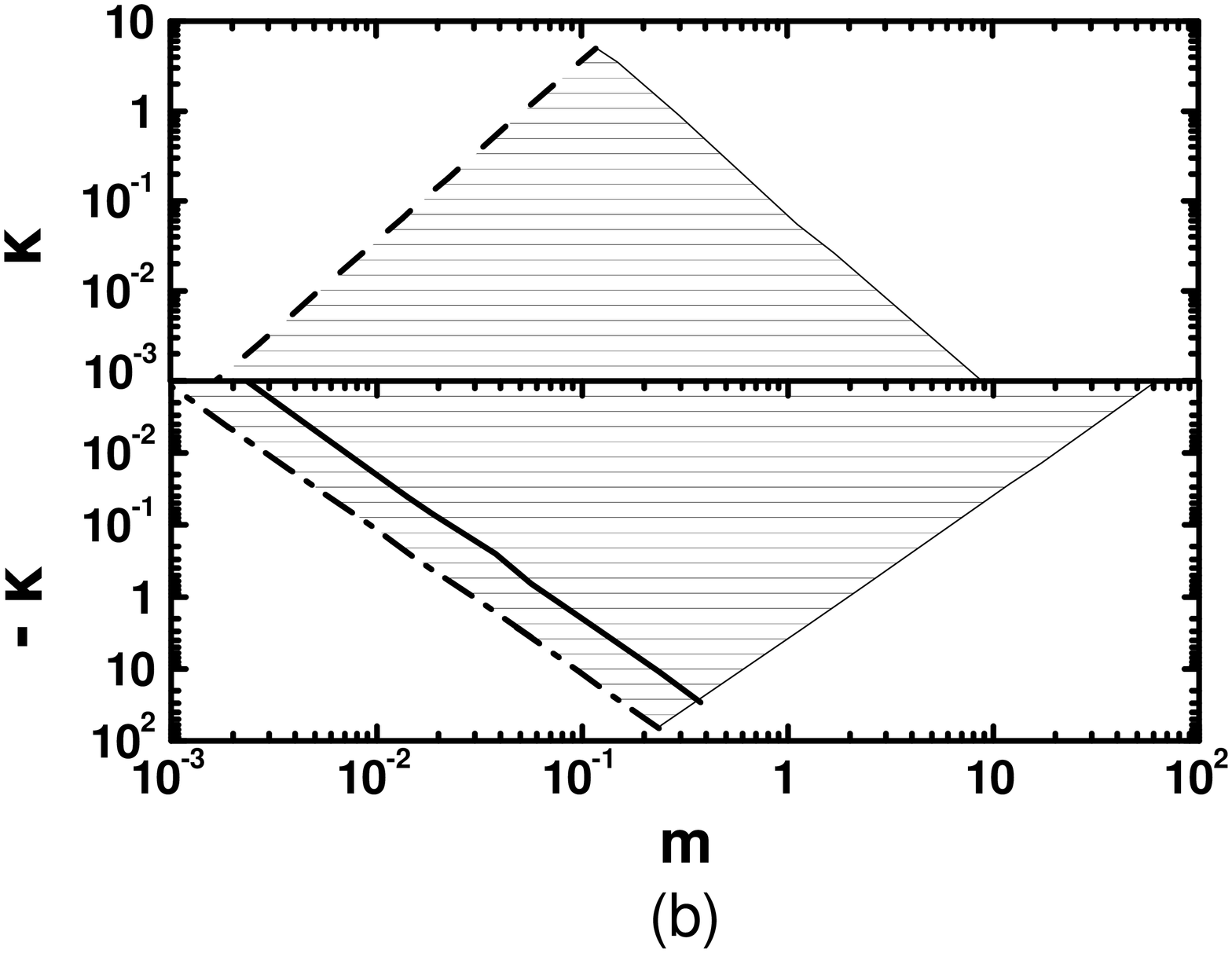,height=3.6in,angle=-90} \hfill
\end{minipage}\vspace*{-.1in}
\begin{center}
\epsfig{file=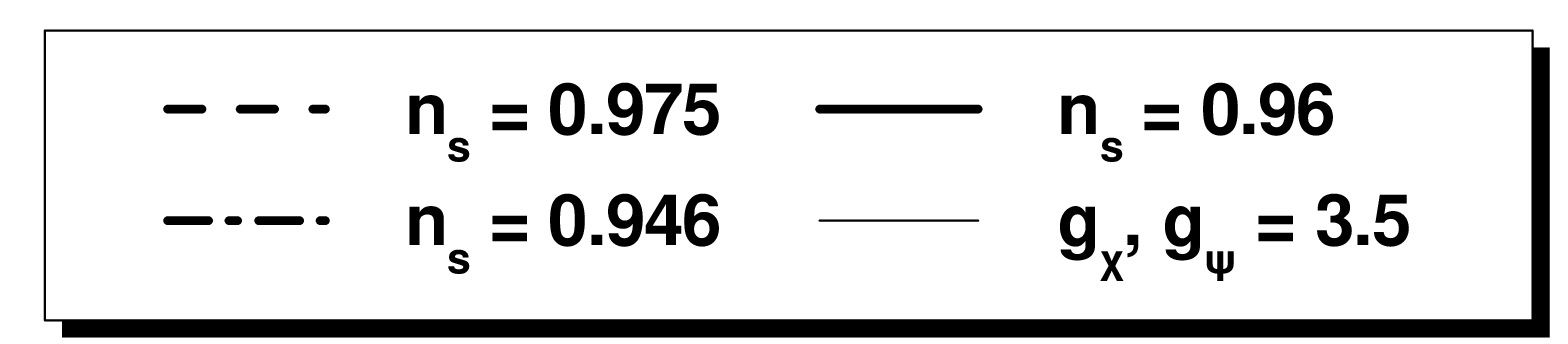,height=2.5in,angle=-90}
\end{center}
\hfill \caption[]{\sl\small Allowed regions (hatched) compatible
with Eqs.~(\ref{Ntot}), (\ref{Prob}), (\ref{nswmap}) and
(\ref{subP}) in the $m-\kp$ plane for $\dV(\sgx)=0$
({\sffamily\ftn a}) or $\dV(\sgf)=0$ ({\sffamily\ftn b}). The
conventions adopted for the various lines are also shown.}
\label{mkn}
\end{figure}

%%%%%%%%%%%%%%%%%%%%%%%%%%%%%%%%%%%%%%%%%%

Varying $m$ and $\kp$ we specify in \Fref{mkn} the available
parameter space from the constraints of \Sref{obs} of the model
for $\Ld$ such that $\dV(\sgx)=0$ (${\sf\ftn a}$) or $\dV(\sgf)=0$
(${\sf\ftn b}$). The conventions adopted for the various lines are
also shown. In particular, the dashed [dot-dashed] lines
correspond to $\ns=0.975$ [$\ns=0.946$], whereas the solid lines
are obtained by fixing $\ns=0.96$ -- see Eq.~(\ref{nswmap}). Along
the thin lines $\gx$ ($\kp>0$) or $\gy$ ($\kp<0$) saturate their
perturbative limit of $\sqrt{4\pi}\simeq3.54$. Obviously, for
$\dV(\sgx)=0$ [$\dV(\sgf)=0$] the allowed region is extended to
roughly larger [lower] $\kp$'s. Focusing on $\ns\simeq0.96$ with
$\sgx\simeq(0.003-1)$ or $|\se_\star|\simeq(0-7)$ and
$\Ns\simeq(54.4-54.8)$ we find
\beqs\bea\label{nmires1}
&0.017\lesssim{\gy}\lesssim3.5,~0.011\lesssim{m/0.1}\lesssim2.4
~~~\mbox{and}~~~77\lesssim\ck\lesssim1.6\cdot10^4
&\mbox{for}~~~\dV(\sgx)=0,\\ \label{nmires2}
 & 0.010\lesssim{\gy}\lesssim3.5,~0.010\lesssim{m/0.1}\lesssim3.5
~~~\mbox{and}~~~77\lesssim\ck\lesssim2.5\cdot10^4&\mbox{for}~~~\dV(\sgf)=0.
~~~~~~~~~~~~~~\eea\eeqs
In both cases $r\simeq0.003$ and $\msn\simeq1.2\cdot10^{-5}$.
Letting $\ns$ vary within the range of \sEref{nswmap}{a} we obtain
$r\simeq(2.4-4.5)\cdot10^{-3}$ and
$\msn\simeq(1-1.4)\cdot10^{-5}$.

Recapitulating this section, we could conclude that the presence
of a large linear non-minimal coupling to gravity by itself or in
conjunction with RCs in the inflationary potential leads to
acceptable results for the observables obtained in the context of
a non-SUSY quadratic inflationary model. The resulting $r$,
though, is well below the sensitivity of the present experiments
\cite{gws, pdust}.

%\newpage
\section{Inflation in SUGRA}\label{fhi}

In this section we move on to the analysis of our SUGRA
realizations of nMI. Namely, in Sec.~\ref{fhi1} we extract the
inflationary potential for any $n$, and we then present our
results for the two radically different cases: taking $n=0$ in
\Sref{fgi} and $n<0$ in \Sref{fgin}.

\subsection{Inflationary Potential}\label{fhi1}

The (tree level) inflationary potential, $\Vhio$, is obtained by
applying \Eref{Vsugra} for $z^\al=\Phi,S$ and $\Whi, \Khi$ given
in \eqs{Whi}{Kol}. If we express $\Phi$ and $S$ according to the
standard parametrization
\beq \Phi=\:{\phi\,e^{i
\th}}/{\sqrt{2}}\>\>\>\mbox{and}\>\>\>S=\:(s +i\bar
s)/\sqrt{2}\,,\label{cannor} \eeq
and confine ourselves along the inflationary track, i.e., for
\beq \th=s=\bar s=0,\label{inftr} \eeq
we find that the only surviving term is
\beqs\beq \Vhio=\Ve(\th=s=\bar s=0)=e^{K}K^{SS^*}\,
|W_{,S}|^2=\frac{m^2|\Phi|^2}{2\fsp\fr^{2+3n}}\,,\label{Vhig}\eeq
where we take into account that
\beq
e^{K}=\fr^{-3(1+n)}\>\>\>\mbox{and}\>\>\>K^{SS^*}={\fr/\fsp}.\label{Vhigg}\eeq
Calculating $\fr$ and $\fsp$ through the expressions
\bea \label{frsp}
\fr=1+\ck{\xsg}-{\kns\xsg^2-\kpp\xsg^4\over6(1+n)}\>\>\>\mbox{and}\>\>\>\fsp=\Omega_{,SS^*}=1-\ksp\xsg\,,\eea\eeqs
and plugging them into \Eref{Vhig}, we find that $\Vhio$ takes the
form
\beq \Vhio=\frac{m^2\sg^{2}}{2\fsp\fr^{2+3n}}
\simeq\frac{m^2\xsg^{2}}{2\fsp} \lf1+\ck\sg\rg^{-(2+3n)}
\simeq\frac{m^2 \xsg^{-3n}}{2\fsp\ck^{2+3n}}\,.\label{3Vhiom}\eeq
The corresponding EF Hubble parameter is
\beq
\Hhi={\Vhio^{1/2}/\sqrt{3}}\simeq{m\xsg^{-3n/2}/\sqrt{6\fsp}\ck^{1+3n/2}}\,.
\label{He}\eeq
Given that $\fsp\ll\fr$ with $\ck\gg1$, $\Vhio$ in \Eref{Vhig} is
roughly proportional to $\xsg^{-3n}$. Besides the inflationary
plateau which emerges for $n=0$ and was studied in \cref{nMCI}, a
chaotic-type potential (bounded from below) is also generated for
$n<0$.

The kinetic terms for the various scalars in \Eref{Saction1} can
be brought into the following form
\beqs\beq \label{K3} K_{\al\bbet}\dot z^\al \dot
z^{*\bbet}=\frac12\lf\dot{\se}^{2}+\dot{\what
\th}^{2}\rg+\frac12\lf\dot{\what s}^2 +\dot{\what{\overline
s}}^2\rg\,.\eeq
Here the dot denotes derivation w.r.t the JF cosmic time and the
hatted fields read
\beq  \label{Jg} {d\widehat \sg\over
d\sg}=\sqrt{K_{\Phi\Phi^*}}=J\simeq{\sqrt{3(1+n)}\over\sqrt{2}\xsg},\>\>\>
\what{\th}= J\,\th\xsg\>\>\>\mbox{and}\>\>\>(\what s,\what{\bar
s})=\sqrt{K_{SS^*}} {(s,\bar s)}\,,\eeq\eeqs
where $K_{SS^*}=\fsp/\fr\simeq1/\ck\xsg$ -- cf. \Eref{Vhigg}. The
spinors $\psi_\Phi$ and $\psi_S$ associated with $S$ and $\Phi$
are normalized similarly, i.e.,
$\what\psi_{S}=\sqrt{K_{SS^*}}\psi_{S}$ and
$\what\psi_{\Phi}=\sqrt{K_{\Phi\Phi^*}}\psi_{\Phi}$. Integrating
the first equation in \Eref{Jg}, we can identify the EF field as
\beq
\se=\sqrt{3(1+n)/2}\ln\sg~~\Rightarrow~~\sg=e^{\sqrt{2/3(1+n)}\se},\label{se1}\eeq
and derive $\Vhio$ as a function of $\se$, i.e.
\beq \label{Vse}\Vhio\simeq
{m^2e^{-\sqrt{6/(1+n)}n\se}}/{2\fsp\ck^{2+3n}}\,.\eeq
From the last expression we can easily infer that, for $n\neq0$,
$\Vhio$ declines away from the so-called $\alpha$-attractor models
\cite{aroest} which are tied to deviations from the conventional
($-3$) coefficient of the logarithm in the \Ka, and the resulting
inflationary potential has the form
$V_0(1-e^{-\sqrt{2/3(1+n)}\se})^2$.

\renewcommand{\arraystretch}{1.4}

\begin{table}[!t]
\bec\begin{tabular}{|c|c|l|}\hline
{\sc Fields} &{\sc Eingestates} & \hspace*{3.cm}{\sc Masses Squared}\\
\hline \hline
$1$ real scalar &$\what{\th}$ & $\what m^2_{\th}\simeq m^2 \ck
(2 + 3 n) \sg^3 /3 (1 + n)\fr^{3 (1 + n)}\simeq4\He_{\rm CI}^2$\\
$2$ real scalars &$\what{s},~\what{\bar s}$ & $\what m^2_{s}= m^2
\sg \lf4-\ck  \sg  \lf 12 n+\ck  (2 -9 n^2 -24\ks (1 + n)\rg \sg
\right.$ \\
&&$\left. + 24 \ck ^2\ks(1 + n) \sg^2)\rg/6(1+n)\ck\fr^{3 (1 + n)} $\\
\hline
$2$ Weyl spinors & $\what{\psi}_\pm={\what{\psi}_{\Phi}\pm
\what{\psi}_{S}\over\sqrt{2}}$& $\what m^2_{\psi\pm}\simeq m^2(2 -
3 \ck  n \sg)^2/6(1 + n)\ck^2\fr^{(2+3 n)} $
\\ \hline
\end{tabular}\ec
\hfill \caption[]{\sl\small Mass spectrum along the trajectory in
\Eref{inftr}.}\label{tab2}
\end{table}

The stability of the configuration in \Eref{inftr} can be checked
by verifying the validity of the conditions
\beqs\beq \left.{\partial
\Ve\over\partial\what\chi^\al}\right|_{\mbox{\Eref{inftr}}}=0\>\>\>
\mbox{and}\>\>\>\what m^2_{
\chi^\al}>0\>\>\>\mbox{with}\>\>\>\chi^\al=\th,s,\bar
s.\label{Vcon} \eeq
Here $\what m^2_{\chi^\al}$ are the eigenvalues of the mass matrix
with elements
\beq \label{wM2}\what
M^2_{\al\bt}=\left.{\partial^2\Ve\over\partial\what\chi^\al\partial\what\chi^\beta}\right|_{\mbox{\Eref{inftr}}}
\mbox{with}\>\>\>\chi^\al=\th,s,\bar s\,,\eeq\eeqs
and hat denotes the EF canonically normalized fields. Upon
diagonalization of $\what M^2_{\al\bt}$ in \Eref{wM2} we can
construct the scalar mass spectrum of the theory along the
direction in \Eref{inftr}. Taking the limits $\kpp\to0, \ksp\to0$
and $\kns\to0$, we find the expressions of the relevant masses
squared, arranged in \Tref{tab2}, which approach rather well the
quite lengthy, exact expressions taken into account in our
numerical computation. As usual -- cf. \cref{nIG, r2} -- the only
dangerous eignestate of $\what M^2_{\al\bt}$ is $\what m^2_{{s}}$
which can become positive and heavy enough by conveniently
selecting $\kx>0$ -- see Secs.~\ref{fgi2} and \ref{fgin2}. Besides
the stability requirement in \Eref{Vcon}, from the derived
spectrum we can numerically verify that the various masses remain
greater than $\Hhi$ during the last $50$ e-foldings of nMI, and so
any inflationary perturbations of the fields other than the
inflaton are safely eliminated. Due to the large effective masses
that $\th,s$ and $\bar s$ in \Eref{wM2} acquire during CI, they
enter a phase of oscillations about zero with decreasing
amplitude. As a consequence, the $\sg$ dependence in their
normalization -- see \Eref{Jg} -- does not affect their dynamics.
Moreover, we can observe that the fermionic (4) and bosonic (4)
degrees of freedom are equal -- here we take into account that
$\what\phi$ is not perturbed.

Inserting the derived mass spectrum in the well-known
Coleman-Weinberg formula \cite{cw}, we find that the one-loop
corrected inflationary potential is
\beq\Vhi=\Vhio +\dV\>\>\>\mbox{with}\>\>\>\dV={1\over64\pi^2}\lf
\widehat m_{\th}^4\ln{\widehat m_{\th}^2\over\Lambda^2} +2
\widehat m_{s}^4\ln{\widehat m_{s}^2\over\Lambda^2}-4\widehat
m_{\psi_{\pm}}^4\ln{\widehat m_{\psi_{\pm}}^2\over\Lambda^2}\rg
,\label{Vhic}\eeq
% +{1\over32\pi^2}\lf \widehat m_{\th}^2
%+2 \widehat m_{s}^2-4\widehat m_{\psi_{\pm}}^2\rg\Lambda^2\\
where $\Lambda$ is a renormalization group  mass scale, $\widehat
m_{\th}$ and $\widehat m_{s}=\widehat m_{\bar s}$ are defined in
\Eref{Vcon} and $\what m_{\psi_{\pm}}$ are the  mass eigenvalues
which correspond to the fermion eigenstates
$\widehat\psi_{\pm}\simeq(\what\psi_{S}\pm\what\psi_{\Phi})/\sqrt{2}$.
Following the strategy adopted in \Sref{nmi}, we determine $\Ld$
by requiring $\dV(\sgx)=0$ or $\dV(\sgf)=0$. Contrary to that
case, though, we ignore here possible contributions to $\Delta V$
from couplings of the inflaton to the lighter degrees of freedom
for two main reasons. First, these couplings are model dependent
-- i.e. they could be non-renormalizable (and so suppressed);
second, in the SUSY framework there are almost identical
contributions to $\dV$ from bosonic and fermionic degrees of
freedom which cancel each other out -- see \cref{r2}. As a
consequence, the possible dependence of our results on the choice
of $\Lambda$ can be totally avoided if we confine ourselves to
$k_S\sim0.5$ for $n=0$ or $k_S\sim0.1$ for $n<0$ resulting to
$\Ld\simeq(2-3)\cdot10^{14}~\GeV$ and
$\Ld\simeq(2-9)\cdot10^{15}~\GeV$ respectively -- see
Secs.~\ref{fgi2} and \ref{fgin2}. Under these circumstances, our
results can be exclusively reproduced by using $\Vhio$.

%Therefore, the predictability of these models is more robust.

\begin{figure}[!t]\vspace*{-.44in}\begin{tabular}[!h]{cc}\begin{minipage}[t]{7.in}
%\begin{center}
\hspace*{.6in}
\epsfig{file=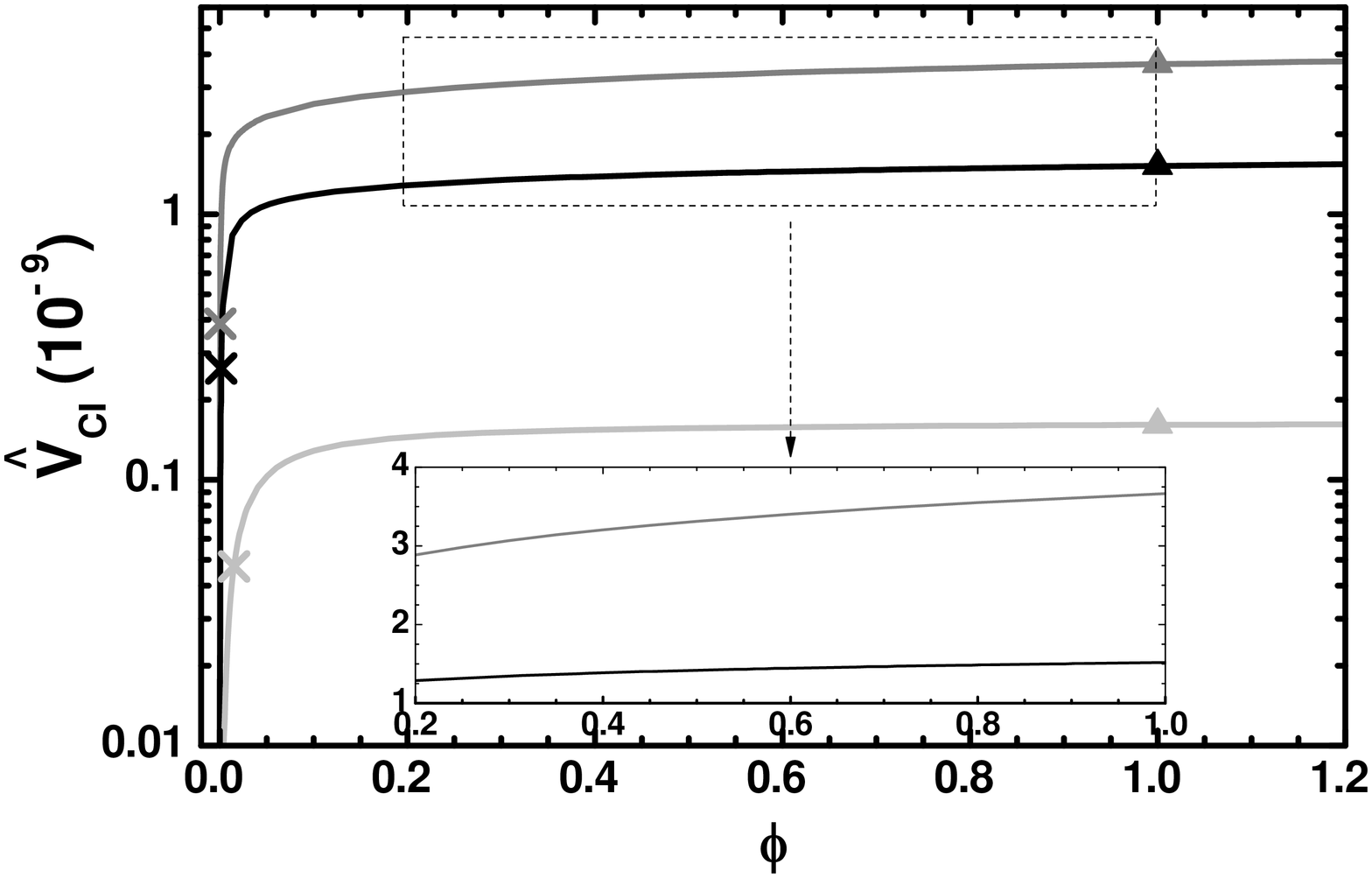,height=3.65in,angle=-90}\end{minipage}
&\begin{minipage}[h]{3.in}
\hspace{-2.5in}{\vspace*{-2.5in}\includegraphics[height=3.5cm,angle=-90]
{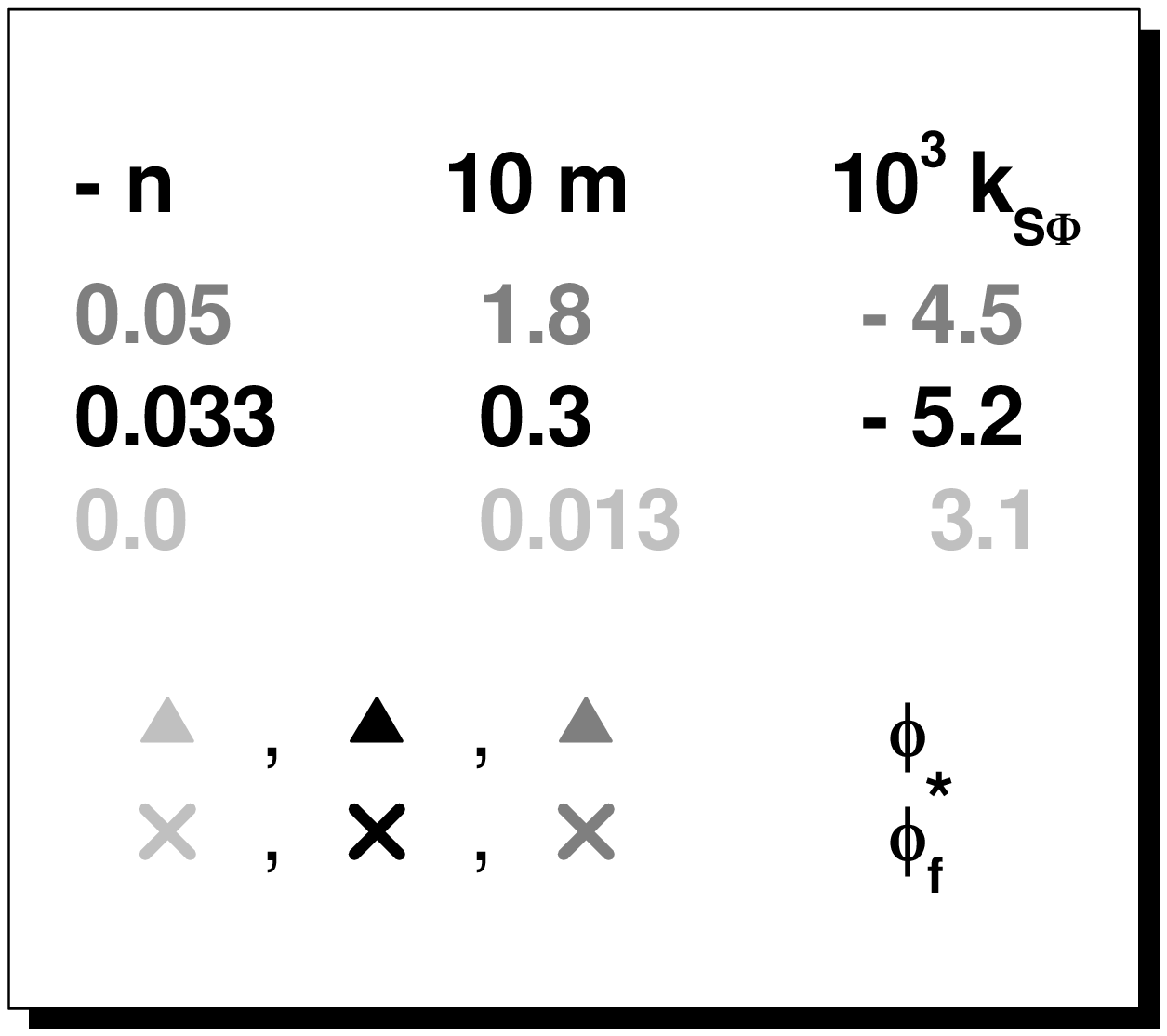}}\end{minipage}
\end{tabular}  \hfill \caption[]{\sl \small Inflationary
potential $\Vhi$ (light gray, black and gray line) as a function
of $\sg$ for $\sg\geq0$, $n=0,-1/30,-1/20$, $m=0.0013,0.03,0.18$
and $\ksp\simeq0.0031,-0.0052,-0.0045$. Values corresponding to
$\sgx$ and $\sgf$ are also depicted.}\label{fig3m}
\end{figure}

The structure of $\Vhi$ as a function of $\sg$ for various $n$'s
is displayed in \Fref{fig3m}, where we depict $\Vhi$ versus $\sg$
imposing $\sgx=1$. The selected values of $m, \ksp$ and $n$, shown
in \Fref{fig3m}, yield $\ns=0.96$ and $r=0.0048, 0.047, 0.11$ for
increasing $|n|$'s -- light gray, black and gray line. The
corresponding $\ck$ values are $(0.77, 7.8, 38.5)\cdot10^2$. We
remark that a gap of about one order of magnitude emerges between
$\Vhio(\sgx)$ for $|n|$ of order $0.01$ and $n=0$ thanks to the
larger $m$ and $\ck$ values; actually, in the former case,
$\Vhio^{1/4}(\sgx)$ approaches the SUSY grand-unification scale,
$8.2\cdot10^{-3}$, which is imperative -- see, e.g.,
\cref{rRiotto} -- for achieving $r$ of order $0.1$. We also
observe that $\Vhio$ close to $\sg=\sgx$ for $n<0$ acquires a
steeper slope which is expected to have an imprint in elevating
$\eph$ -- see \Sref{fgin} -- and, via \sEref{ns}{c}, on $r$.

%%%%%%%%%%%%%%%%%%%%%%%%%%%%%%%%%%%%%%%%%%%%%%%%%%%%%%%%%%%%%%%%%%%%
%%%%%%%%%%%%%%%%%%%%%%%%%%%%%%%%%%%%%%%%%

\subsection{$n=0$ Case}\label{fgi}

We focus first on the form of \Ka\ induced by \Eref{Kol} with
$n=0$. Our analysis in \Sref{fgi1} presents some approximate
expressions which assist us to interpret the numerical results
exhibited in \Sref{fgi2}.

\subsubsection{Analytic Results.}\label{fgi1}

Upon substitution of \eqs{3Vhiom}{Jg} into \Eref{sr}, we can
extract the slow-roll parameters during the inflationary stage.
Namely, we find
\beq \eph=\frac43\lf\frac1{\ck\sg}  + \ksp
\sg^2\rg^2\>\>\>\mbox{and}\>\>\> \ith=\frac{8 - 4 \ck  \sg  (1 - 2
\ksp \sg^2 (5 + \ck  \sg ))}{3 \fr^2 \fsp^2}\, \cdot\label{gsr1}
\eeq
It can be numerically verified that $\sgx, \sgf$ and $\Ns$ do not
decline a lot from their values for $\ksp=0$. Therefore,
\Eref{res2} stabilizes our scheme against higher order terms in
$\fk$ -- see \Eref{Frdef} -- despite the fact that \Eref{se11}
yields $\se>1$ even for $\sgx<1$. Also, \Eref{nmiN} can serve for
our estimates below. In particular, replacing $\Vhio$ from
\Eref{3Vhiom} and $\sgx$ from \Eref{nmiN} in \Eref{Prob} we obtain
\begin{equation} \sqrt{\As}=\frac{m \sgx\sqrt{1 - \ksp \sgx^2}}{8 (\pi + \ck  \ksp \pi
 \sgx^3)}\>\>\Rightarrow\>\> m=\lf27 \ck^2  + 64 \ksp  \Ns^3\rg\frac{2\pi\sqrt{\As}}{9 \Ns\ck},\label{lang} \eeq
which is quite similar to \Eref{Prob2} obtained in the non-SUSY
case. Inserting \Eref{nmiN} into \eqs{gsr1}{ns} and expanding for
$\ck\gg1$, we extract the following expressions for the
observables
\beq \ns\simeq1 - \frac{2}{\Ns} + \frac{256}{27}\frac{\ksp
\Ns^2}{\ck
^2},\>\>\>\as\simeq-\frac{2}{\Ns^2}-\frac{640}{27}\frac{\ksp
\Ns}{\ck ^2}\>\>\>\mbox{and}\>\>\>r\simeq\frac{12}{\Ns^2} +
\frac{512}{9}\frac{\ksp \Ns}{\ck ^2}\cdot \label{rns}\eeq
As in the case of \Eref{nmins}, the emergent depedence of the
observables on $\ksp$ is stronger for $\ns$, since it goes as
$\Ns^2/\ck^2$, and weaker for $\as$ and $r$, since they go as
$\Ns/\ck^2$. This depedence does not exist within no-scale SUGRA
since $\ksp$ vanishes by definition -- see \Eref{nsks}.

\subsubsection{Numerical Results.}\label{fgi2}

The present inflationary scenario depends on the parameters:
\beq
m,\>\ck,\>\kx,\>\ksp,\>\kpp\>\>\>\mbox{and}\>\>\>\Trh\,.\label{para1}\eeq
Our results are essentially independent of $\kx$ values, provided
that $\what m_{s}^2>0$ for every allowed $m$ and $\ck$ -- see
\Tref{tab2}. The same is also valid for $\kpp$ since the
contribution from the second term in $\fr$, \Eref{frsp}, is
overshadowed by the strong enough first term including $\ck\gg1$.
We therefore set $\kx=\kpp=0.5$. Since we do not specify the
interaction of the inflaton to the light degrees of freedom,
$\Trh$ is a free parameter. We choose $\Trh=4.1\cdot10^{-10}$
which is a typical value encountered in similar settings -- cf.
\cref{nMCI,nmH,r2}. Besides these values, in our numerical code,
we use as input parameters $\ck,~\ksp$ and $\sgx$. For every
chosen $\ck$, we restrict $m$ and $\sgx$ so that
\eqss{Ntot}{Prob}{subP} are satisfied. In addition, by adjusting
$\ksp$ we can achieve $\ns$ values in the range of
Eq.~(\ref{nswmap}). Our results are displayed in \sFref{fig2g}{a}
[\sFref{fig2g}{b}], where we delineate the hatched regions allowed
by the above restrictions in the $m-\ck$ [$m-\ksp$] plane. We
follow the conventions adopted for the thick lines in \Fref{mkn}.
Along the solid thin line, which provides the lower bound for the
regions presented in \Fref{fig2g}, the constraint of
\sEref{subP}{b} is saturated. At the other end, the allowed
regions terminate along the faint dashed line where $|\ksp|=3$,
since we expect $\ksp$ values of order unity to be natural.

%We limit ourselves to this upper bound on $\ksp$

From \sFref{fig2g}{a} we see that $\ck$ remains almost
proportional to $m$ and for constant $m$, $\ck$ increases as $\ns$
decreases. From \sFref{fig2g}{b} we note that $\ksp$ takes natural
(order unity) values for $1.94\lesssim m/0.01\lesssim7.8$ or
$6\gtrsim\sgx/0.1\gtrsim0.15$. For lower $m$ or larger $\sgx$
values some degree of tuning ($\sim0.01$) is needed since $\ksp$
is confined close to zero for $n_{\rm s}=0.96$, whereas for larger
$m$ or lower $\sgx$ values, $\ksp$ starts increasing sharply
beyond unity. More explicitly, for $\ns=0.96$ and
$\Ne_\star\simeq52$, taking $\sgx=(0.01-1)$ or
$|\se_\star|=(0-8.6)$, we find:
\beq\label{resg} 77\lesssim
\ck\lesssim1.5\cdot10^5\>\>\>\mbox{with}\>\>\>0.49\lesssim
m/0.01\lesssim11.7\>\>\>\mbox{and}\>\>\> 0.0031\lesssim
|\ksp|\lesssim3\,. \eeq
For this range of values, we obtain  $6.8\lesssim
{|\as|/10^{-4}}\lesssim8.2$ and $r\simeq3.8\cdot 10^{-3}$ which
lie within the ranges of \Eref{nswmap}. On the other hand, the
results within no-scale SUGRA are much more robust since the
$\ksp$ (and $\kpp$) dependence collapses -- see \Eref{nsks}.
Indeed, no-scale SUGRA predicts
$\ns\simeq0.964,~\as=-6.5\cdot10^{-4}$ and $r=4\cdot10^{-3}$
identically with the non-SUSY case -- see columns E and F of
\Tref{tab1}. The same results would have been achieved, if we had
considered $S$ as a nilpotent superfield \cite{nil} since the
stabilization term $|S|^4$ would have been absent in \Eref{Kol}
and so all the fourth order terms could be avoided.

%%%%%%%%%%%%%%%%%%%%%%%%%%%%%%%%%%%%%%%%%%%%%%%%%%%%%%%%%%%%%%%%%%%%%
\begin{figure}[!t]\vspace*{-.12in}
\hspace*{-.19in}
\begin{minipage}{8in}
\epsfig{file=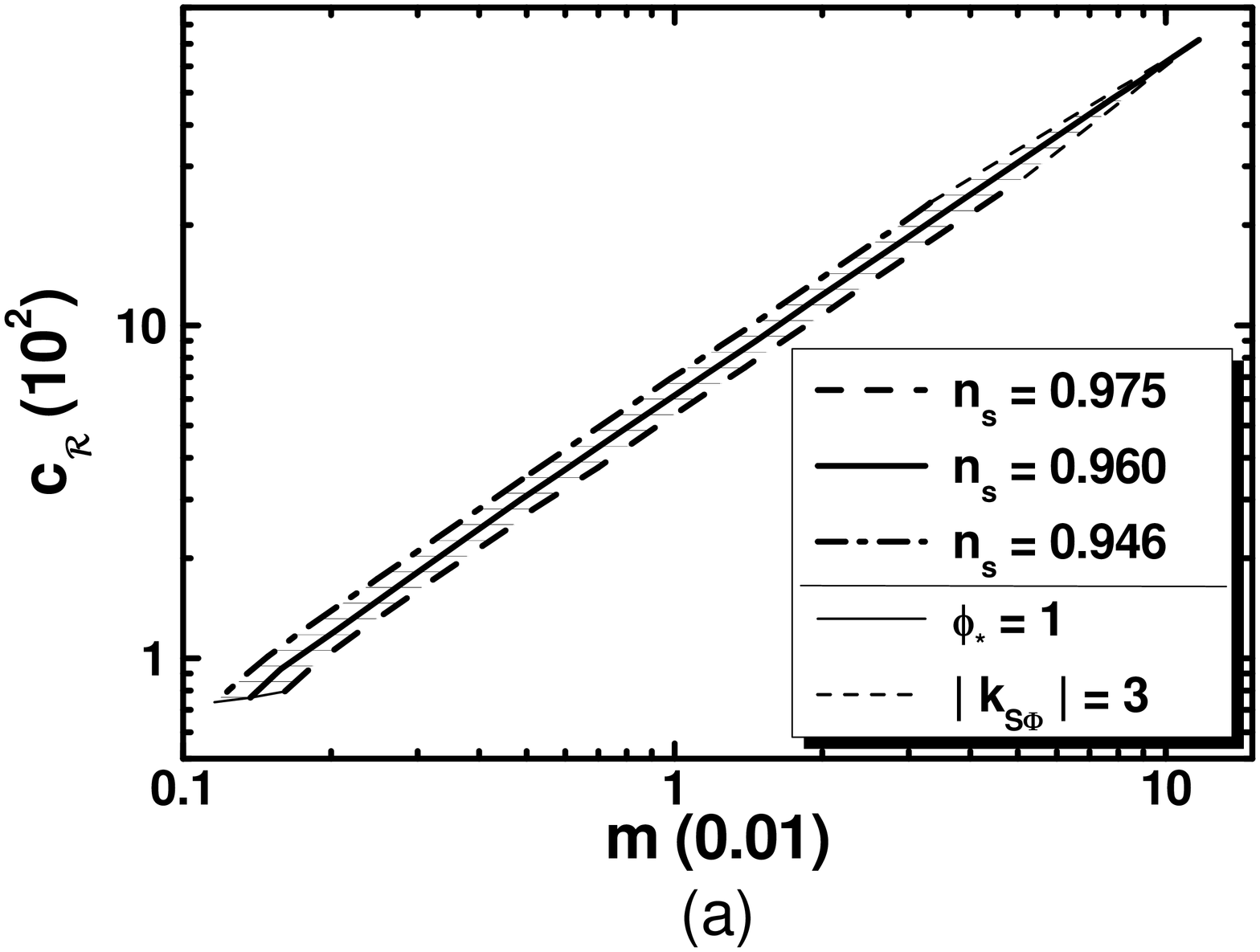,height=3.6in,angle=-90}
\hspace*{-1.2cm}
\epsfig{file=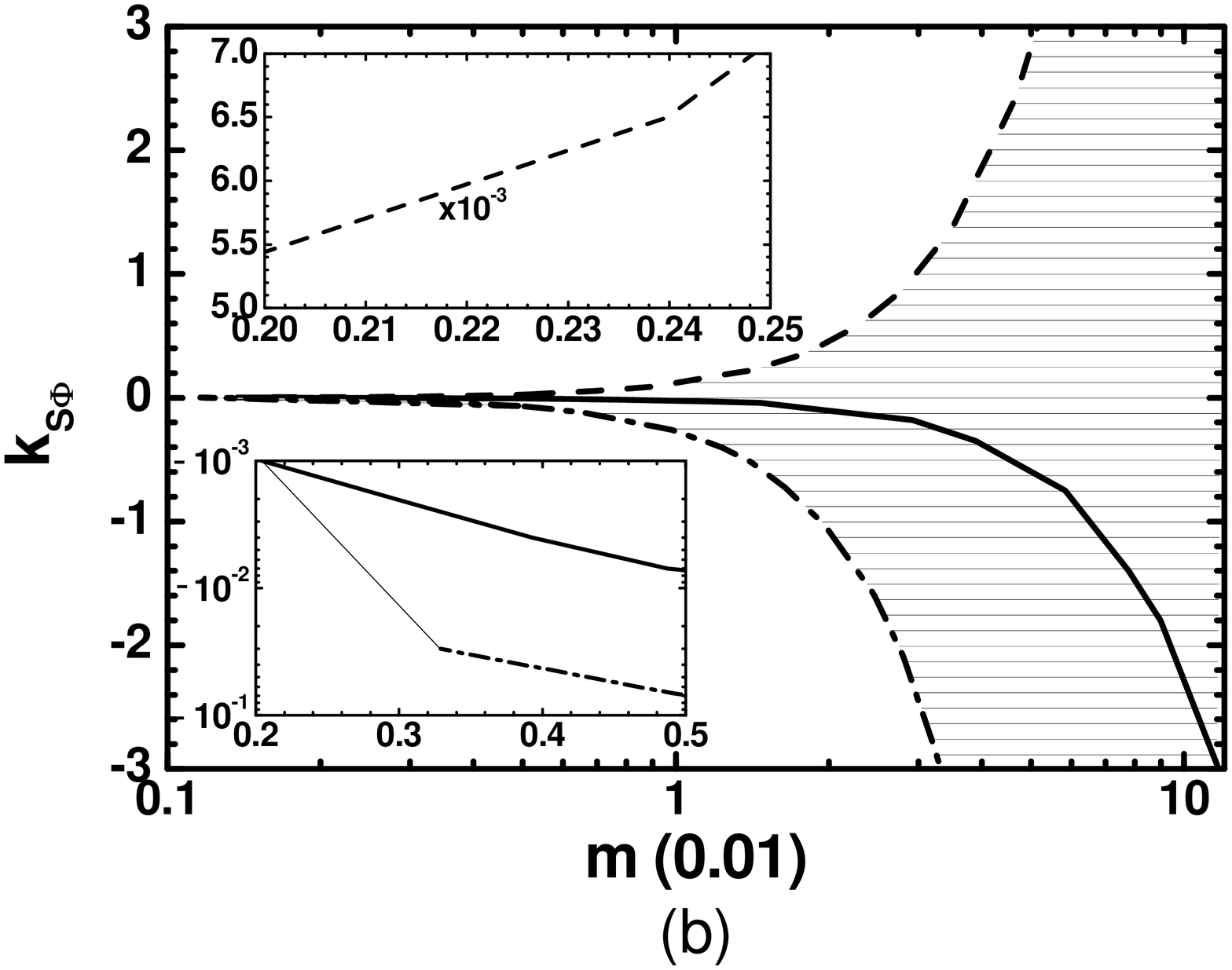,height=3.6in,angle=-90} \hfill
\end{minipage}
\hfill \caption[]{\sl\small Allowed regions (hatched) compatible
with Eqs.~(\ref{Ntot}), (\ref{Prob}), (\ref{nswmap}) and
(\ref{subP}) in the $m-\ck$ ({\sffamily\ftn a}) and $m-\ksp$
({\sffamily\ftn b}) plane for $n=0$ and $\ks=\kpp=0.5$. The
conventions adopted for the various lines are shown in panel
{\sffamily\ftn (a)}.}\label{fig2g}
\end{figure}

%%%%%%%%%%%%%%%%%%%%%%%%%%%%%%%%%%%%%%%%%%

\subsection{$n<0$ Case}\label{fgin}

Following the strategy of the previous section, we present below
first some analytic results in \Sref{fgin1}, which provides a
taste of the numerical findings exhibited in \Sref{fgin2}.

\subsubsection{Analytic Results.}\label{fgin1}

Plugging \eqs{3Vhiom}{Jg} into \Eref{sr}, we obtain the following
approximate expressions for the slow-roll parameters
\beqs\beq \eph=\frac{(2 - 3 \ck  n \sg  + \ck  \ksp (2 + 3 n)
\sg^3)^2}{3 (1 + n) \fr^2\fsp^2}\label{gsrn1} \eeq and
\beq\ith=\frac{8-2 \ck \sg (2+15 n -2 \ksp (10 \sg^2+ 2 \ck \ksp
(2-6n)\sg^3)}{3 (1 + n) \fr^2\fsp^2}\,\cdot\label{gsrn2} \eeq\eeqs
Taking the limit of the expressions above for $\ksp\simeq0$, we
can analytically solve the condition in \Eref{srcon} w.r.t $\xsg$.
The results are
\beq \sg_{1\rm f} =
\frac{2}{\sqrt{3}(\sqrt{3}n+\sqrt{1+n})\ck}\>\>\>\mbox{and}\>\>\>
\sg_{2\rm f} = \frac{8}{(2 + 15 n + \sqrt{28 + 84 n + 81 n^2})\ck
}\cdot\label{sgf}\eeq
The termination of nMI mostly occurs at $\sgf=\sg_{1\rm f}$
because we mainly get $\sg_{1\rm f}>\sg_{2\rm f}$.

Given that $\sgf\ll\sgx$ we can estimate $\Ns$ through \Eref{Nhi},
\beqs\beq  \Ns=\frac32 (1 + n)\lf\frac{\ln\sgx}{2} -\frac{(2+3 n)
\ln(2 - 3 \ck n \sgx)}{6 n}\rg\cdot\label{Ngm}\eeq
Neglecting the first term in the last equality and solving w.r.t
$\sgx$, we get an indicative value for $\sgx$
\beq\label{sm*}\sgx=\lf2- \re\rg/3n\ck,\>\>\>\mbox{with}\>\>\>\re
= e^{-4 n \Ns/(1 + n) (2 + 3 n)}\,.\eeq
Although a radically different dependence of $\sgx$ on $\Ns$
arises -- cf. \Eref{nmiN} -- $\sgx$ can again remain \sub\ for
large $\ck$'s fulfilling \sEref{subP}{b}. Indeed,
\beq \label{fmsub} \sgx\leq1\>\>\>\Rightarrow\>\>\>\ck\geq
(2-\re)/3n\,.\eeq\eeqs
As in the previous cases -- see Secs.~\ref{resnmi2} and \ref{fgi1}
-- $\se$  corresponding to $\sgx$ and $\sgf$ turn out to be \trns,
since plugging \eqs{sm*}{sgf} into \Eref{se1} we find
\beq \se_\star\simeq\sqrt{3(1 + n)\over2}\lf {4|n|\Ns\over(1 +
n)(2+3n)}-\ln 3|n|\ck \rg\>\>\>\mbox{and}\>\>\>\se_{\rm
f}\simeq\sqrt{3(1 + n)\over2}\ln
{2/\sqrt{3}\over(\sqrt{3}n+\sqrt{1+n})\ck}\,, \label{sme*}\eeq
which give $|\se_\star|\simeq(0.5-4)$ and $|\se_{\rm
f}|\simeq(7.7-13.4)$ for $n=-(0.03-0.05)$ -- in rather good
agreement with the numerical results which yield
$|\se_\star|\simeq(0-3.8)$ and identical results for $|\se_{\rm
f}|$. Despite this fact, our construction remains stable since the
dangerous higher order terms are exclusively expressed as
functions of the initial field $\Phi$, and remain harmless for
$|\Phi|\leq1$.

Upon substitution of \Eref{sm*} into \Eref{Prob} we end up with
\begin{equation} m \simeq\frac{4 \sqrt{\As} \pi (1 + (2 -\re)/3 n)^{3 n/2}
(27 n^3 \re\ck ^2  - \ksp (2 + 3 n) (\re-2)^3)}{3 n \sqrt{1 + n}
(\re-2) \sqrt{9  n^2 \ck ^2-\ksp (\re-2)^2}}\cdot \label{langm}
\eeq
We remark that $m$ remains almost proportional to $\ck$ -- cf.
~\Eref{lang} -- but it also depends both on $\ksp$ and $n$.
Inserting \Eref{sm*} into \eqs{gsrn1}{gsrn2}, then employing
\sEref{ns}{a} and expanding for $\ck\gg1$, we find
\beqs\beq \label{nsgm} \ns=1-2 n\frac{8-4\re+3n(4+\re(\re-2))}{(1
+ n) (\re- 3 n - 2 )^2}+ \frac{8 \ksp (2 + 3 n) (\re-2)^3}{27
n^2(1 + n)(\re-3 n-2)\ck ^2}\cdot \eeq
Following the same steps, from \sEref{ns}{c} we find
\beq \label{ragm} r=16\lf\frac{3n^2\re^2}{(1 + n) (\re-3 n-2)^2} -
4\ksp  \frac{(\re - 2)^2\re}{9n (1 + n)(\re-3 n-
2)\ck^2}\rg\cdot\eeq\eeqs
From the above expressions we see that primarily $|n|\neq0$ and
secondarily $n<0$ help to reduce $\ns$ below unity and sizably
increase $r$. On the other hand, the dependence of $r$ on $\ksp$
is rather weak since the dominant contribution originates from the
first term, which is independent of $\ksp$ and $\ck$, whereas the
correction from the second term is suppressed by an inverse power
of $\ck^2$. On the contrary, the depedence of $\ns$ on $\ksp<0$ is
somehow stronger since the presence of $\ck^2\gg1$ in the
denominator of the second term is accompanied by the factor
$n^2\ll1$, which compensates for the reduction of the
corresponding contribution.

\subsubsection{Numerical Results.}\label{fgin2}

Besides the free parameters shown in \Eref{para1} we also have $n$
which is constrained to negative values. Using the reasoning
explained in \Sref{fgi2} we set $\kpp=0.5$ and
$\Trh=4.1\cdot10^{-10}$. On the other hand, $\what m_{s}^2$ can
become positive with $\ks$ lower than the value used in
\Sref{fgi2} since positive contributions from $n<0$ arises here --
see \Tref{tab2}. Moreover, if $\ks$ takes a value of order unity
$\what m_{s}^2$ grows more efficiently than in the case with
$n=0$, rendering thereby the RCs in \Eref{Vhic} sizeable for very
large $\ck$ values ($\sim10^5$). To avoid such dependence of the
model predictions on the RCs, we use $\kx$ values to lower than
those used in \Sref{fgi2}. Thus, we set $\kx=0.1$ throughout. As
in the previous case, \eqss{Ntot}{Prob}{subP} assist us to
restrict $m$ (or $\ck\geq1$) and $\sgx$. By adjusting $n$ and
$\ksp$ we can achieve not only $\ns, \as$ and $r$ values in the
range of \Eref{nswmap} but also $r$ values in the observable
region $(0.01-0.1)$.

Confronting the parameters with Eqs.~(\ref{Ntot}), (\ref{Prob}),
(\ref{nswmap}{\sffamily\ftn a, b}) and (\ref{subP}) we depict the
allowed (hatched) regions in the $m-\ck$, $m-\ksp$, $m-r$ and
$m-\as$ planes for $n=-1/30$ (light gray lines and horizontally
hatched regions), $n=-1/25$ (black lines and horizontally hatched
regions), $n=-1/20$ (gray lines and vertically hatched regions) in
\sFref{fig2gm}{a}, {\sf\ftn (b), (c)} and {\sf\ftn (d)}
respectively. In the horizontally hatched regions $r$ is
compatible with \sEref{nswmap}{c}, whereas in the vertically
hatched region $r$ overpasses slightly this bound for
$\ns\gtrsim0.946$. Note that the conventions adopted for the
various lines are identical with those used in \Fref{fig2g} --
i.e., the dashed, solid (thick) and dot-dashed lines correspond to
$\ns=0.975, 0.96$ and $0.946$ respectively, whereas along the thin
(solid) lines the constraint of \sEref{subP}{b} is saturated. The
bound $|\ksp|=3$ limits the various regions at the other end along
the thin dashed line

From \sFref{fig2gm}{a} we remark that $\ck$ remains almost
proportional to $m$ but the dependence on $\ksp$ is weaker than
that shown in \sFref{fig2g}{a}. Also, as $|n|$ increases, the
allowed areas are displaced to larger $m$ and $\ck$ values in
agreement with \Eref{fmsub} -- cf. \Fref{fig2g}. Similarly, the
allowed $\ksp$ values move to the right for increasing $m$ values
and fixed $\ns$ as shown in \sFref{fig2gm}{b}. Indeed, if we
increase $\ck$, \Eref{nsgm} dictates an increase of $\ksp$ in
order to keep $\ns$ constant. This effect deviates somewhat from
our findings in the similar settings of \cref{nIG}. Finally, from
\sFref{fig2gm}{c} and {\sf\ftn (d)} we conclude that employing
$|n|\gtrsim0.01$, $r$ and $\as$ increase w.r.t their values for
$n=0$ -- see results below \Eref{resg}. As a consequence,  for
$n=-0.033$ and $-0.04$, $r$ enters the observable region. An
increase in $r$ even larger than the present bound in
\sEref{nswmap}{c} is also possible. On the other hand, $\as$
although one order larger than its value for $n=0$ remains
sufficiently low; it is thus consistent with the fitting of data
with the standard $\Lambda$CDM model -- see \Eref{nswmap}. As
anticipated below \Eref{ragm}, the resulting $r$ values depend
only on the input $n$ and $\ksp$ (or $\ns$), and are independent
of $m$ (or $\ck$). The same behavior is also true for $\as$. It is
worth noticing that a decrease of $\ksp$ below zero is imperative
in order to simultaneous fulfill Eqs.~(\ref{nswmap}{\sf\ftn a})
and ({\sf\ftn c}). Indeed, had we increased the prefactor $(-3)$
in \Eref{Kol} by eliminating the fourth order terms -- by
assuming, e.g., that $S$ is a nilpotent superfield \cite{nil} --
the enhancement of $r$ would be accompanied with an increase of
$\ns$ which would have become incompatible with \sEref{nswmap}{a}.

%%%%%%%%%%%%%%%%%%%%%%%%%%%%%%%%%%%%%%%%%%%%%%%%%%%%%%%%%%%%%%%%%%%%%
\begin{figure}[!t]\vspace*{-.12in}
\hspace*{-.19in}
\begin{minipage}{8in}
\epsfig{file=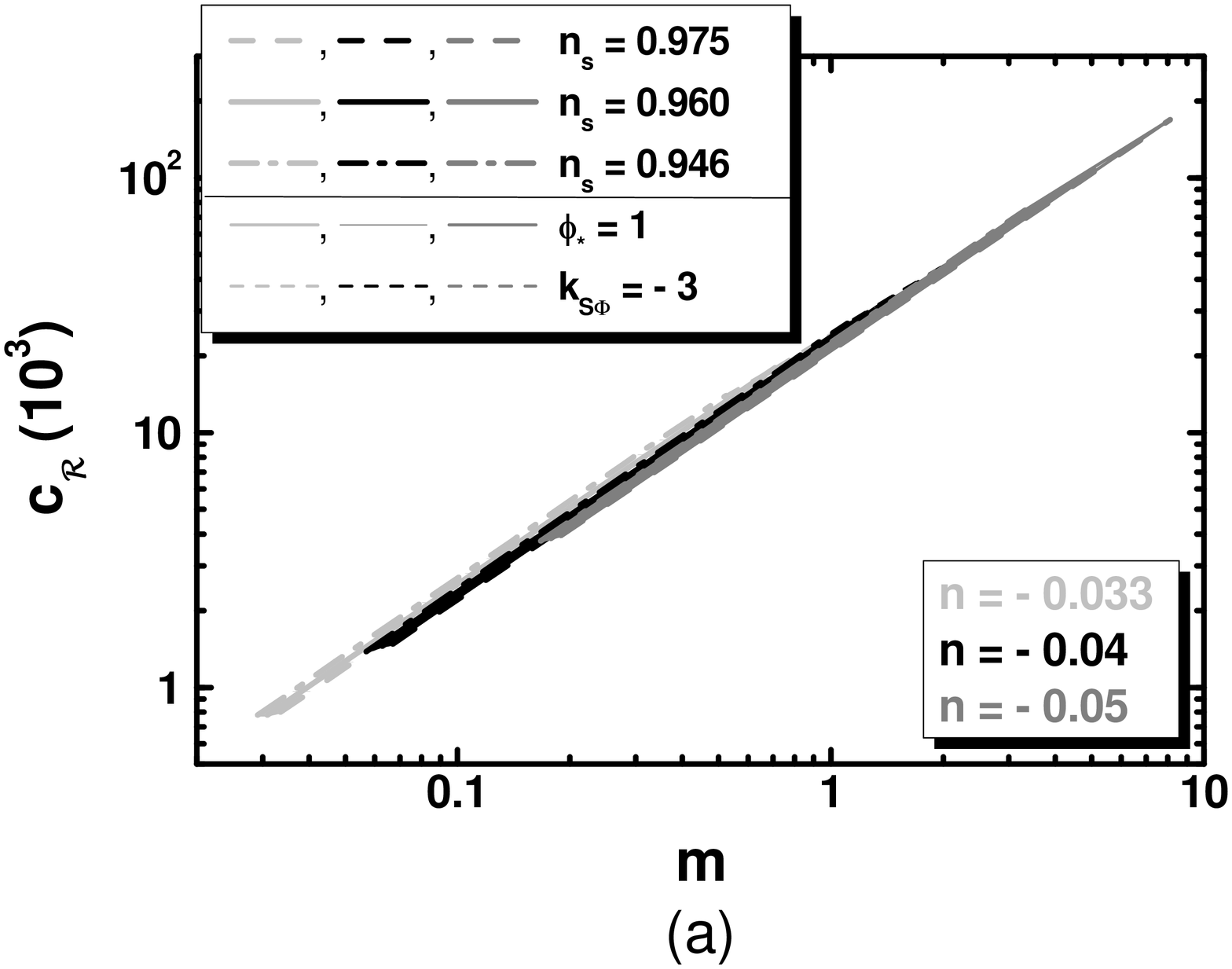,height=3.6in,angle=-90}
\hspace*{-1.2cm}
\epsfig{file=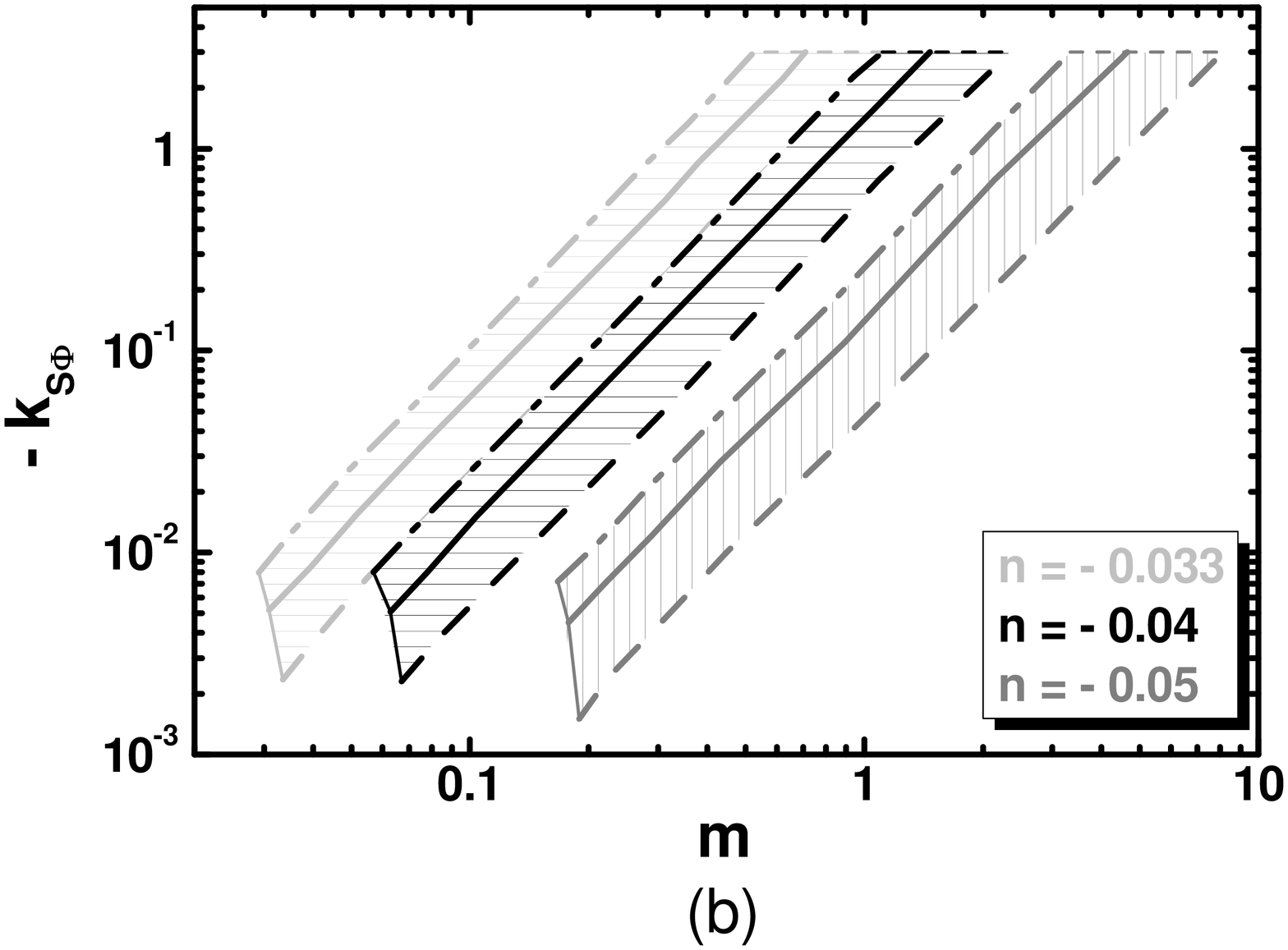,height=3.6in,angle=-90} \hfill
\end{minipage}
\hfill \hspace*{-.19in}
\begin{minipage}{8in}
\epsfig{file=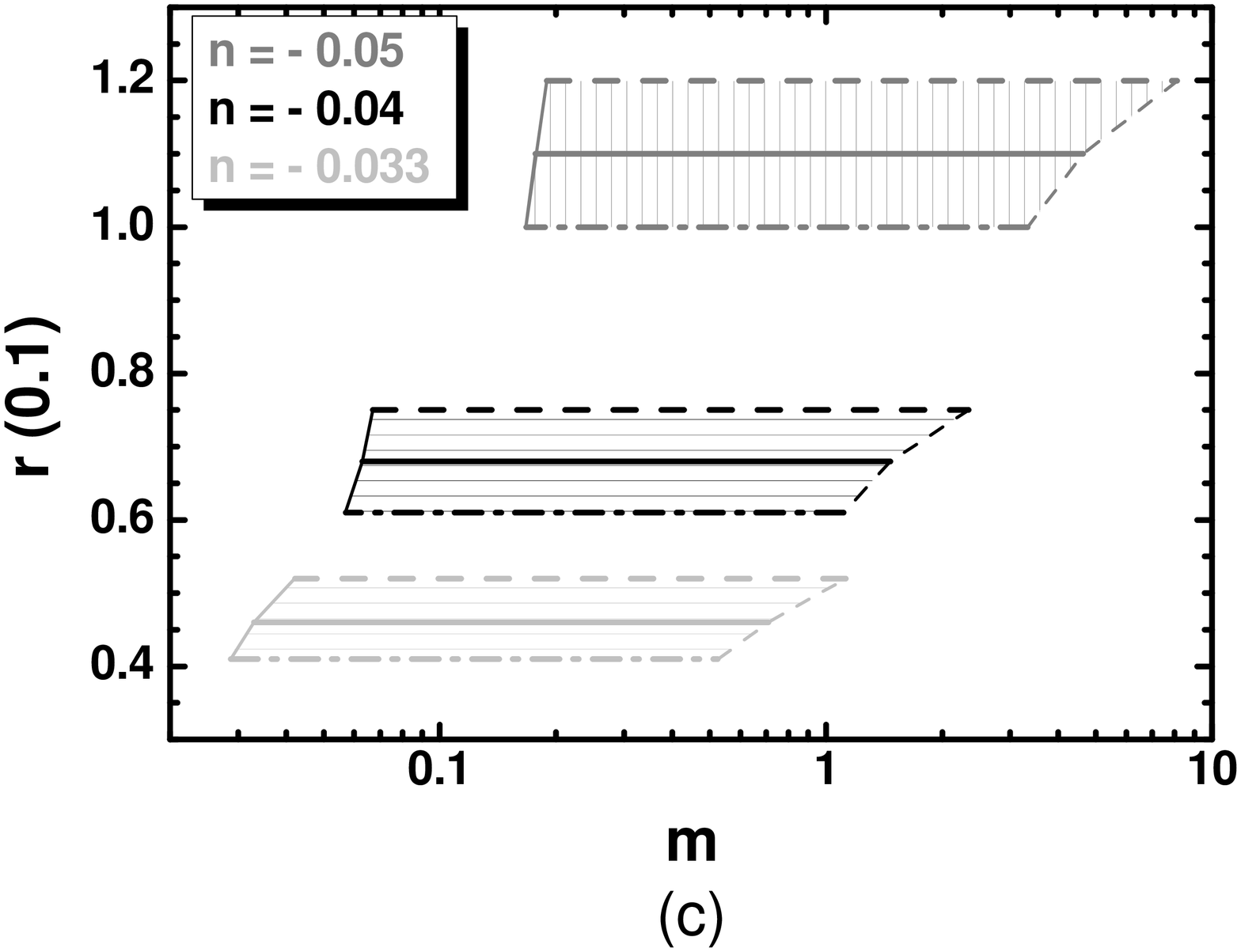,height=3.6in,angle=-90}
\hspace*{-1.2cm}
\epsfig{file=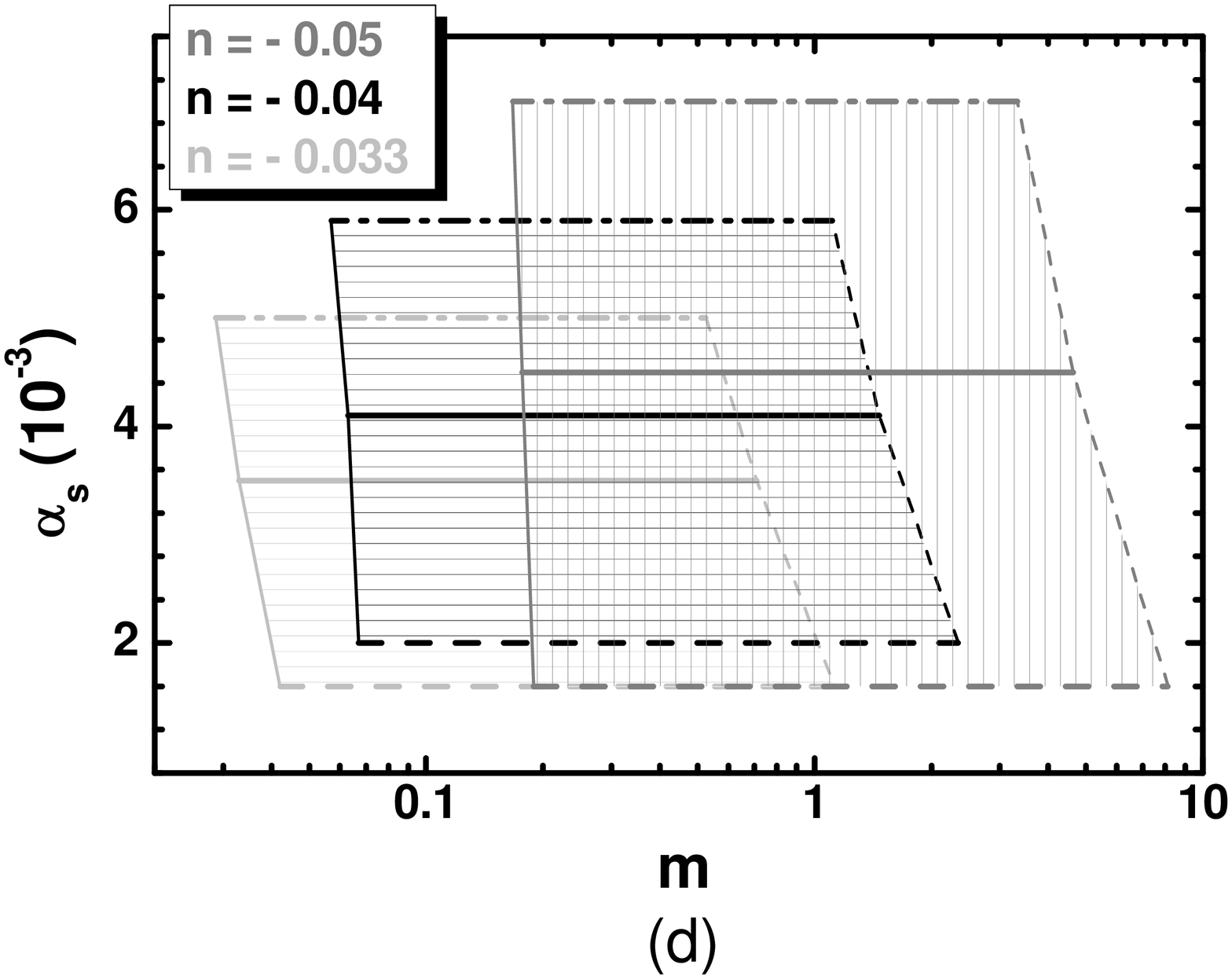,height=3.6in,angle=-90} \hfill
\end{minipage}
\hfill \caption[]{\sl\small Allowed regions (hatched) compatible
with Eqs.~(\ref{Ntot}), (\ref{Prob}), (\ref{nswmap}{\sffamily\ftn
a, b}) and (\ref{subP}) in the $m-\ck$ ({\sffamily\ftn a}),
$m-\ksp$ ({\sffamily\ftn b}), $m-r$ ({\sffamily\ftn c}), $m-\as$
({\sffamily\ftn d}) plane for $\ks=0.1$, $\kpp=0.5$ and $n=-0.033$
(light gray lines and hatched regions), $n=-0.04$ (black lines and
hatched regions), $n=-0.05$ (gray lines and hatched regions). The
conventions adopted for the type and color of the various lines
are shown in panel {\sffamily\ftn (a)}.}\label{fig2gm}
\end{figure}

%%%%%%%%%%%%%%%%%%%%%%%%%%%%%%%%%%%%%%%%%%

More explicitly, for $n_{\rm s}=0.96$ and $\Ne_\star\simeq51.7$ we
find:
\beqs\baq\label{resgm} && 0.78\lesssim {\ck/
10^3}\lesssim18\>\>\>\mbox{with}\>\>\>0.03\lesssim
m\lesssim0.71\>\>\>\mbox{and}\>\>\> 0.005\lesssim
-{\ksp}\lesssim3\, \>\>\>(n=-0.033);\>\>\>\>\>\>\\ &&
\label{resgm2} 1.45\lesssim {\ck/
10^3}\lesssim34\>\>\>\mbox{with}\>\>\>0.06\lesssim
m\lesssim1.47\>\>\>\mbox{and}\>\>\> 0.002\lesssim
-{\ksp}\lesssim3\,\>\>\>(n=-0.04);\>\>\>\>\>\>\>\>\>\>\>\>\>\\
&& \label{resgm3}3.85\lesssim {\ck/
10^3}\lesssim10^2\>\>\>\mbox{with}\>\>\>0.18\lesssim
m\lesssim4.64\>\>\>\mbox{and}\>\>\> 0.0045\lesssim
-{\ksp}\lesssim3\,\>\>\>(n=-0.05).\>\>\>\>\>\>\>\>\>\>\>\>\>
\eaq\eeqs
In these regions, $\sgx$ ranges from $1$ to about $0.04$ and the
remaining observables are
\beq \label{resgm4}  {r\over0.1}=0.46, 0.68,
1.1\>\>\>\mbox{and}\>\>\>
{\as\over0.001}=3.6,4,4.5\>\>\>\mbox{for}\>\>\>
-{n\over0.01}=3.3,4,5\eeq respectively. As in the similar model of
\cref{nIG}, the observable $r$ values above are achieved with
\sub\ $\phi$ values.  Note that this requirement sets strict upper
bound on $r$ \cite{rRiotto, rlimit} -- e.g., in the case of SUSY
hybrid inflation \cite{rfhi} we have $r\leq0.01$. However, in our
present set-up, $\phi$ does not coincide with the EF inflaton,
$\se$, which remains \trns\ and close to the values shown in
\Eref{sme*}. Thus, our results do not contradict the Lyth bound
\cite{lyth}, which applies to $\se$.

%%%%%%%%%%%%%%%%%%%%%%%%%%%%%%%%%%%%%%%%%%%%%%%%%%%%%%%%%%%%%%%%%%%%%
\begin{figure}[!t]\vspace*{-.12in}
\hspace*{-.19in}
\begin{minipage}{8in}
\epsfig{file=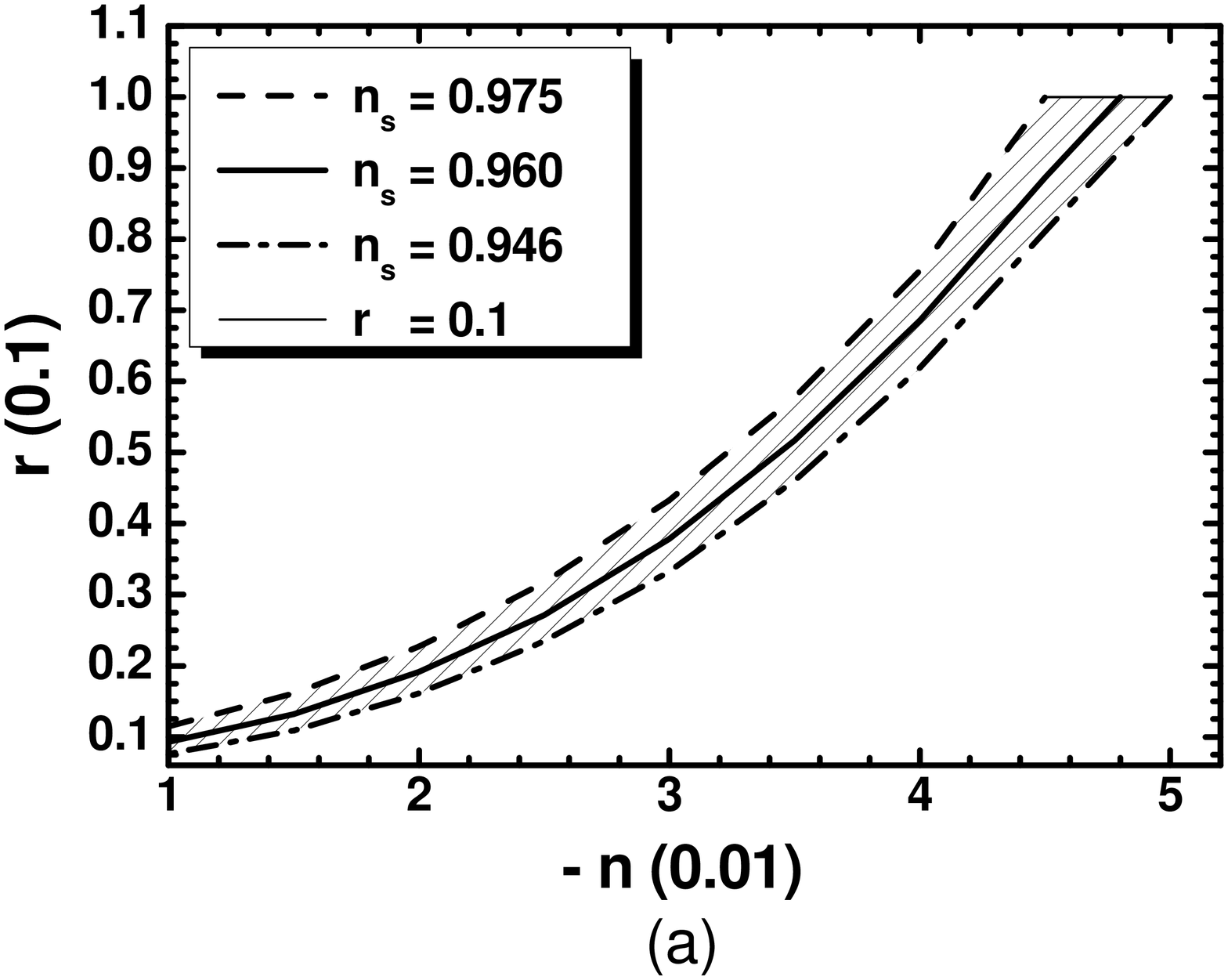,height=3.6in,angle=-90}
\hspace*{-1.2cm}
\epsfig{file=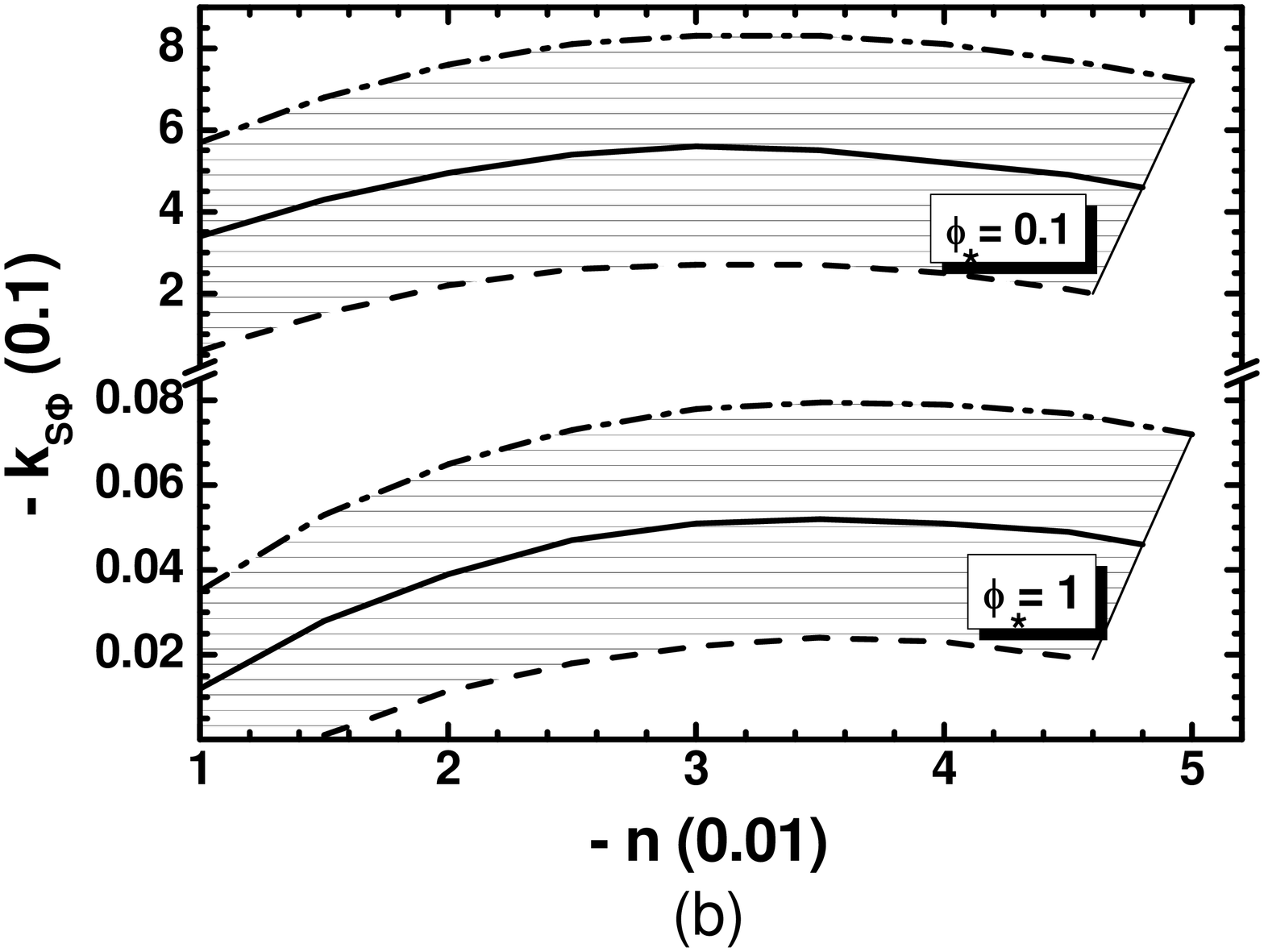,height=3.6in,angle=-90} \hfill
\end{minipage}
\hfill \caption[]{\sl\small Allowed regions (hatched) compatible
with Eqs.~(\ref{Ntot}), (\ref{Prob}), (\ref{nswmap}{\sffamily\ftn
a, b, c}) and (\ref{subP}) in the $(-n)-r$ plane ({\sffamily\ftn
a}) and $(-n)-(-\ksp)$ plane for $\sgx=0.1$ and $1$
({\sffamily\ftn b}). In both panels we set $\ks=0.1$ and
$\kpp=0.5$. The conventions adopted for the various lines are
shown in panel {\sffamily\ftn (a)}.}\label{fign}
\end{figure}

%%%%%%%%%%%%%%%%%%%%%%%%%%%%%%%%%%%%%%%%%%

Taking advantage of the independence of $r$ on $m$ and $\ck$,
highlighted in \sFref{fig2gm}{c}, we can delineate the allowed
region of our model using $n$ as a free parameter. More
specifically, fixing $\sgx$ and $\ns$, we can vary $n$ below zero
to obtain a continuous variation of the derived $r$. This way we
construct the (hatched) regions allowed by all the constraints of
\Sref{obs} in the $(-n)-r$ plane -- see \sFref{fign}{a}. We see
that confining $|n|$ in the range $(0.01-0.05)$ comfortably
assures observable $r$ values for any $\ns$ in the range of
\sEref{nswmap}{a}. We use the same shape code for the the various
thick lines as in Fig.~\ref{fig2g} and \ref{fig2gm}, whereas along
the thin line here \sEref{nswmap}{c} is saturated. In
\sFref{fign}{b} we display the allowed regions in the
$(-n)-(-\ksp)$ plane for $\sgx=0.1$ (upper island) and $1$ (lower
island). In all, for $n_{\rm s}=0.96, r=(0.01-0.1)$ and
$\Ne_\star\simeq51.7$ we take:
\beqs\bea\label{resgmr} && 1.32\lesssim {\ck\over
10^2}\lesssim31\>\>\>\mbox{with}\>\>\>0.03\lesssim
m/0.1\lesssim1.4\>\>\>\mbox{and}\>\>\> 1.2\lesssim
-{\ksp/0.001}\lesssim4.6\, \>\>\>(\sgx=1);\>\>\>\>\>\>\>\>\>\>\>\\
&& \label{resgmr2} 1.32\lesssim {\ck\over
10^3}\lesssim31\>\>\>\mbox{with}\>\>\>0.03\lesssim
m\lesssim1.42\>\>\>\mbox{and}\>\>\> 3.4\lesssim
-{\ksp/0.1}\lesssim4.6\,\>\>\>(\sgx=0.1);\>\>\>\>\>
 \eea\eeqs
From these results we infer that $\ksp$ takes more natural (order
one) values for lower $\sgx$ values. In fact, from \Eref{sm*} we
deduce that $\sgx$ decreases as $\ck$ increases and \Eref{nsgm}
entails an augmentation of $\ksp$ in order the $\ns$ value to be
kept unchanged.

\section{Effective Cut-off Scale}\label{fhi3}

An outstanding trademark of nMI with linear coupling to gravity is
that it is unitarity-safe, despite the fact that its
implementation with \sub\ $\phi$ values -- see \eqs{res2}{fmsub}
-- requires relatively large $\ck$ values. To show that this fact
-- first noticed in \cref{riotto} -- is valid for all our models
we  extract below the UV cut-off scale, $\Qef$, expanding the
action in \Eref{action1} in the JF -- see \Sref{fhi3b} -- or this
in \Eref{action} in the EF -- see \Sref{fhi3a}. Although the
expansions about $\vev{\phi}=0$, presented below, are not valid
\cite{cutof} during nMI, we consider the $\Qef$ extracted this way
as the overall cut-off scale of the theory, since the reheating
phase -- realized via oscillations about $\vev{\sg}$ -- is an
unavoidable stage of the inflationary dynamics.

\subsection{Jordan Frame Computation}\label{fhi3b}

Thanks to the special dependence of $\fr$ on $\phi$ there is no
interaction between the excitation of $\phi$ about $\vev{\phi}=0$,
$\dph$, and the graviton, $h^{\mu\nu}$ which can jeopardize the
validity of perturbative unitarity. Indeed, expanding $g_{\mu\nu}$
about the flat spacetime metric $\eta_{\mu\nu}$ and the inflaton
$\phi$ about its v.e.v,
\beq
g_{\mu\nu}\simeq\eta_{\mu\nu}+h_{\mu\nu}\>\>\>\mbox{and}\>\>\>\phi=0
+\dph\,,\eeq
and retaining only the terms with two derivatives of the
excitations, the part of the lagrangian corresponding to the two
first terms in the right-hand side of \Eref{action} takes the form
\cite{cutof, nIG}
\beqs\baq \nonumber \delta{\cal L}&=&-{\vev{\fr}\over8}{F}_{\rm
EH}\lf h^{\mu\nu}\rg +\frac12\vev{F_{\rm K}}\partial_\mu
\dph\partial^\mu\dph+\frac12F_{\cal R}\lf \vev{f_{\cal
R,\phi}}\dph+\frac12 \vev{f_{\cal R,\phi\phi}}\dph^2\rg\\
 &=&-{1\over8}F_{\rm EH}\lf \bar h^{\mu\nu}\rg+
\frac12\partial_\mu
\overline\dph\partial^\mu\overline\dph,\label{L2}\eaq
where $F_{\rm K}=1$ for the non-SUSY case and $F_{\rm
K}=\kns+3n\ck^2/2$ for our SUGRA scenaria; the functions $F_{\rm
EH}$ and $F_{\cal R}$ are given in \cref{nIG}; $\bar h_{\mu\nu}$
and $\overline\dph$ are the JF canonically normalized fields
defined by the relations
\beq \overline\dph=\sqrt{\frac{\vev{\fr}}{\vev{\bar f_{\cal
R}}}}\dph\>\>\>\mbox{and}\>\>\> \bar h_{\mu\nu}=
\sqrt{\vev{\fr}}\,h_{\mu\nu}+\frac{\vev{f_{\cal
R,\phi}}}{\sqrt{\vev{\fr}}}\eta_{\mu\nu}\dph
\>\>\>\mbox{with}\>\>\>\bar f_{\cal R}=F_{\rm K}\fr+\frac32
f_{\cal R,\phi}^2\,,\label{Jcan}\eeq\eeqs
where $\vev{\fr}=1$ and $\vev{\bar f_{\cal R}}=F_{\rm
K}+3\ck^2/2$. Since $f_{\cal R,\phi\phi}$ vanishes in the non-SUSY
regime -- see \sEref{Vci}{b} -- and is independent on $\ck$ in our
SUSY scenario -- see \Eref{frsp} --, no interaction $\dph^2h$ with
$h=h^\mu_\mu$ appears -- cf. \cref{cutoff,cutof,nIG} -- and so the
theory does not face any problem with the perturbative unitarity.

\subsection{Einstein Frame Computation}\label{fhi3a}

Alternatively, $\Qef$ can be determined in EF, following the
systematic approach of \cref{riotto}. We concentrate here on the
SUGRA version of our model. The result for the non-SUSY case can
be easily recovered for $n=0$ and $\kns=1$. Note, in passing, that
the EF (canonically normalized) inflaton, in the SUGRA version is
\beq\dphi=\vev{J}\dph\>\>\>\mbox{with}\>\>\>\vev{J}\simeq\sqrt{3(1+n)\ck^2/2+\kns}\,,
\label{dphi} \eeq
and it acquires mass given by \Eref{msn1}. Comparing \Eref{dphi}
with \Eref{dphi1}, we infer that the results are identical with
the non-SUSY case only for the no-scale scenario. Making use of
\Eref{lang}, we find $\msn=1.3\cdot10^{-5}$ for the no-scale SUGRA
case. Beyond no-scale SUGRA with $n=0$, replacing $m$ in
\Eref{msn1} from \Eref{lang}, we find that $\msn$ inherits from
$m$ a mild dependence on $\ksp$. E.g., for $n=0$ and $\ns$ in the
range of \sEref{nswmap}{a} we find
$1.2\lesssim\msn/10^{-5}\lesssim1.5$ with the lower [upper] value
corresponding to the lower [upper] bound on $\ns$ in
\Eref{nswmap}. For the same $\ns$, when $n<0$ and $r=(0.01-0.1)$
we get, using \Eref{langm}, $2\lesssim\msn/10^{-5}\lesssim4$
independently of the $\sgx$ value.

The fact that $\dphi$ does not coincide with $\dph$ -- contrary to
the standard Higgs nMI \cite{cutoff,cutof} -- ensures that our
models are valid up to $\mP$ -- from now on we restore the
presence of $\mP$ in the formulas. To show it, we write the EF
action ${\sf S}$ in \Eref{Saction1} along the path of \Eref{inftr}
as follows
\beqs \beq\label{S3} {\sf S}=\int d^4x \sqrt{-\what{
\mathfrak{g}}}\lf-\frac{1}{2} \rce +\frac12\,J^2
\dot\phi^2-\Ve_{\rm CI0}+\cdots\rg, \eeq
where the ellipsis represents terms irrelevant for our analysis;
$J$ and $\Vhio$ are given by \eqs{Jg}{3Vhiom} respectively.
Expanding $J^2 \dot\phi^2$ about $\vev{\phi}=0$ in terms of
$\dphi$ in \Eref{dphi} we arrive at the following result,
\beq\label{exp2} J^2
\dot\phi^2=\lf1-\frac{2}{\sqrt{1+n}}\sqrt{\frac{2}{3}}\frac{\dphi}{\mP}+\frac{2}{(1+n)}\frac{\dphi^2}{\mP^2}-
\frac{8}{3(1+n)^{3/2}}\sqrt{\frac{2}{3}}\frac{\dphi^3}{\mP^3}+{20\over9(1+n)^2}\frac{\dphi^4}{\mP^4}-\cdots\rg\dot\dphi^2.\eeq
On the other hand, $\Vhio$ in \Eref{3Vhiom} can be expanded about
$\vev{\phi}$ as follows
\beq\Vhio=\frac{m^2\dphi^2}{3(1+n)\ck^2}\lf1-\frac{\sqrt{2}(2+3n)}{\sqrt{3(1+n)}}\frac{\dphi}{\mP}+(2+3n)\frac{\dphi^2}{\mP^2}
-\sqrt{\frac{2}{3}}\frac{8+9n(2+n)}{3\sqrt{1+n}}\frac{\dphi^3}{\mP^3}+\cdots\rg\cdot\label{Vexp}\eeq\eeqs
From the expressions above, \eqs{exp2}{Vexp}, we can infer that
$\Qef=\mP$.

%\newpage
\section{Conclusions}\label{con}

Inspired by the recently released \plk\ results \cite{pdust} on
the dust polarization data which appear to refute the original
interpretation \cite{gws} of the \bicep\ data in terms of
$r\simeq(0.16-0.2)$, we have explored quadratic CI accompanied by
a strong enough linear non-minimal coupling $\fr\gg1$ of the
inflaton to gravity. Imposing a lower bound on $\ck$ involved in
$\fr$, we succeeded to realize nMI for \sub\ values of the
inflaton, stabilizing thereby our predictions from possible higher
order corrections. Moreover, in all cases, the corresponding
effective theory is valid up to the Planck scale.

In the non-SUSY context we investigated ramifications to the
initially proposed scenario \cite{nmi} due to the presence of RCs
generated by the coupling of the inflaton to matter. We showed
that the RCs can affect the results on $\ns$ but not $r$ which
remains at the presently unobservable level. For $\ns=0.96$ the
model favors fermionic coupling of the inflaton with strength in
the range $(0.01-3.5)$ and predicts $r\simeq0.003$ and
$\msn\simeq3\cdot10^{13}~\GeV$.

In the SUSY framework, we considered a superpotential with a
bilinear term including two chiral superfields, which leads to the
quadratic potential and can be uniquely determined by imposing an
$R$ symmetry and a global $U(1)$ symmetry. On the other hand, we
extended our analysis in \cref{nMCI} by considering a quite
generic form of logarithmic \Ka. Namely, the prefactor multiplying
the logarithm was formulated as $(-3)(1+n)$, and all possible
terms up to the fourth order in powers of the various fields were
considered apart from the one needed to cure the tachyonic
instability, occurring along the direction of the accompanying
non-inflaton field -- see \Eref{Kol}.

In the case of no-scale SUGRA, thanks to the underlying
symmetries, the inflaton is not mixed with the accompanying
non-inflaton field in the \Ka. As a consequence, the model
predicts $\ns\simeq0.963$, $\as\simeq-0.00065$ and $r\simeq0.004$,
in excellent agreement with the current \plk\ data, and
$\msn\simeq3\cdot10^{13}~\GeV$. Beyond no-scale SUGRA, for $n=0$,
we showed that $\ns$ spans the entire allowed range in
\sEref{nswmap}{a} by conveniently adjusting the coefficient
$\ksp$. In addition, for $n\simeq-(0.01-0.05)$, $r$ becomes
accessible to the ongoing measurements with negligibly small
$\as$. In this last case a mild tuning of $\kx$ to values of order
$0.1$ is adequate so that the one-loop RCs remain subdominant and
$\msn$ is confined to the range $(5-9)\cdot10^{13}~\GeV$.

To conclude, the presence of a strong linear non-minimal coupling
of the inflation to gravity can comfortably rescue CI based on the
quadratic potential in both the non-SUSY and the SUSY context.

\acknowledgments{We would like to thank  F. Bezrukov, A. Lahanas,
G. Lazarides, J.~Papavassiliou, M.~Postma, A.~Santamaria and
O.~Vives for useful discussions. C.P.~acknowledges support from
the Generalitat Valenciana under grant PROMETEOII/2013/017 and
Q.S. from the DOE grant No. DE-FG02-12ER41808.}

\appendix{Imaginary Quadratic Inflation}

\renewenvironment{subequations}{%
\refstepcounter{equation}%
% \theparentequation{\theequation}%
\setcounter{parentequation}{\value{equation}}%
  \setcounter{equation}{0}
  \def\theequation{A.\theparentequation{\sf\ftn \alph{equation}}}%
  \ignorespaces
}{%
  \setcounter{equation}{\value{parentequation}}%
  \ignorespacesafterend
}

\paragraph{} The \Ka\ used in the no-scale
version of our model -- see \eqs{Kol}{nsks} -- exclusively depends
on the combination
\beq F_{\rm s}=\Phi+\Phi^*,\label{Fshift}\eeq
which enjoys a shift symmetry w.r.t the inflaton superfield
$\Phi$, according to which
\beq \Phi \to\ \Phi+ic,\label{shift}\eeq
where $c$ is a real number -- cf.~\cref{shift1}. Consequently, a
tantalizing question, which should be addressed is whether CI can
be realized though the imaginary part of $\Phi$, as advocated in
\cref{rStar} for the Starobinsky model -- for another approach to
the same problem see \cref{rEllis}. To be more specific, we check
below if the combination of the superpotential in \Eref{Whi} with
the following \Ka
\beq  \Khi=-3(1+n)\ln\lf 1+{\ck\over \sqrt{2}}\,F_{\rm s}-{
|S|^2+\kpp F_{\rm s}^2+2\ksp |S|^2F_{\rm s}-\bksp|S|^2 F_{\rm s}^2
-\kx|S|^4\over3(1+n)}\rg\,,\label{Kim}\eeq
supports inflationary solutions. To analyze this possibility, we
find it convenient to decompose the fields into real and imaginary
parts as follows
\beq \Phi=\:{(\bar\phi +i
\phi)}/{\sqrt{2}}\>\>\>\mbox{and}\>\>\>S=\:{(\bar s +i
s)}/{\sqrt{2}}\,.\label{cannorim} \eeq
Thanks to the shift symmetry in \Eref{shift}, $K$ is independent
of $\phi$ and the direction
\beq \bar\phi=s=\bar s=0,\label{inftrim} \eeq
is a good candidate inflationary trajectory. Along it we find that
$\Ve$ in \Eref{Vsugra}, with $W$ and $K$ given in \eqs{Whi}{Kim},
takes the form of the purely quadratic potential, i.e.,
\beq \label{Vcim}\Vhio=\Ve\lf\bar\phi=s=\bar s=0\rg=\frac12
m^2\phi^2.\eeq
The kinetic terms of the various scalars in \Eref{Saction1} can be
brought into the following form
\beqs\beq \label{K3im} K_{\al\bbet}\dot z^\al \dot
z^{*\bbet}=\frac12\lf\dot{\se}^{2}+\dot{\what{\overline\phi}}^{2}\rg+\frac12\lf\dot{\what
s}^2 +\dot{\what{\overline s}}^2\rg,\eeq
where the hatted fields are defined as
\beq  \label{Jim} {d\widehat \sg\over
d\sg}=\sqrt{K_{\Phi\Phi^*}}=J=\sqrt{2 \kpp + 3(1 + n) \ck^2
/2},\>\>\> \what{\bar{\phi}}=
J\,\bar{\phi}\>\>\>\mbox{and}\>\>\>(\what s,\what{\bar
s})={(s,\bar s)}.\eeq\eeqs
The corresponding hatted spinors are normalized similarly, i.e.,
$\what\psi_{S}=\psi_{S}$ and $\what\psi_{\Phi}=J\psi_{\Phi}$.
However, the trajectory in \Eref{inftrim} is not generally stable
w.r.t the direction $\bar\phi$. This can be rectified if we impose
the condition:
\beq \left.\frac{\partial
\Ve}{\partial\what{\overline{\phi}}}\right|_{\mbox{\Eref{inftrim}}}=-\frac1{2J}
m^2 \lf 2 \sqrt{2} \ksp + (2 + 3 n)\ck\rg \sg^2=0~~\Rightarrow~~
\ksp = -(2 + 3 n)\ck/2\sqrt{2},\label{Vconim}\eeq
which signals a serious tuning of the parameters. In other words,
the term including $\ksp$ in \Eref{Kim} is imperative for the
validity of this inflationary set-up. Its analysis can be realized
along the lines of \Sref{obs} using the corrected inflationary
potential shown in \Eref{Vhic} if we employ the particle spectrum
displayed in \Tref{tab3}. From there we note that, as in previous
cases, there is an instability along the $s$ and $\bar s$
directions which is avoided if we set $\kx=1$. On the other hand,
$\kpp=\bksp=1$ assures the positivity and heaviness of $\what
m_{\bar\phi}^2$. Moreover, positive $n$ values assist us to obtain
enough e-foldings of CI consistently with \sEref{subP}{b}.

\begin{table}[!t]
\bec\begin{tabular}{|c|c|l|}\hline
{\sc Fields} &{\sc Eingestates} & \hspace*{3.cm}{\sc Masses Squared}\\
\hline \hline
$1$ real scalar &$\what{\bar\phi}$ &$\what m^2_{\bar\phi}= \lf m^2
(6(1 + n) +12\bksp(1+n)\sg^2+ (2 + 3 n) \right.$ \\
&&$\left. (4 \kpp +9 (1 + n)^2\ck^2 ) \sg^2)\rg/6 (1 + n) J^2$\\
$2$ real scalars &$\what{s},~\what{\bar s}$ & $\what m^2_{s}=  m^2
\lf1/J^2 + (6 \kx (1 + n)-1) \sg^2/3(1 + n)\rg$\\ \hline
$2$ Weyl spinors & $\what{\psi}_\pm={\what{\psi}_{\Phi}\pm
\what{\psi}_{S}\over\sqrt{2}}$& $\what m_{\psi\pm}=m/J$
\\ \hline
\end{tabular}\ec
\hfill \caption[]{\sl\small Mass spectrum along the trajectory in
\Eref{inftrim}.}\label{tab3}
\end{table}

Given that the RCs remain subdominant we can approach, quite
precisely, the inflationary dynamics using $\Vhio$ in \Eref{Vcim}.
In particular, applying \eqs{srcon}{Nhi} we find
\beq \what\epsilon(\sgf)=\what\eta(\sgf)=\frac{2}{J^2
\sgf^2}=1~~{\Rightarrow}~~\sgf = \frac{\sqrt{2}}{J}
\>\>\>\mbox{and}\>\>\>\Ns\simeq\frac{J^2 \sgx^2}{4}
~~\Rightarrow~~ \sgx=\frac2J \sqrt{\Ns}\,.\label{im1}\eeq From the
last equality we see that CI can be implemented with \sub\ $\sg$
values for $J\geq2\sqrt{\Ns}$. Taking into account the
normalization in \Eref{Prob} and employing \Eref{im1} we find
\beq \sqrt{\As}=\frac{J m \sgx^2}{4 \pi\sqrt{6}} ~~{\Rightarrow}
~~m = \frac{\sqrt{6\As} J \pi}{\Ns}\cdot\label{mJ}\eeq
That is, contrary to the simplest quadratic CI, $m$ is not
constant (for constant $\Ns$) but proportional to $J$. Upon
substitution of the last expression in \Eref{im1} into \Eref{ns}
we obtain the inflationary observables
\beq \label{obsim}\ns = 1-
{2}/{\Ns}\simeq0.962,\>\>\>\as=-{2}/{\Ns^2}\simeq7.1\cdot10^{-4}
\>\>\>\mbox{and}\>\>\>r ={8}/{\Ns}\simeq0.15\eeq for $\Ns=52.8$
corresponding to $\Trh=4.1\cdot10^{-10}$. As regards the mass of
the inflaton at the vacuum, it can be obtained by inserting
\eqs{Vcim}{Jim} into \Eref{msn1} with result
\beq \label{msnim}\msn =  \sqrt{6\As}
\pi/\Ns\simeq6.7\cdot10^{-6}\,. \eeq
Therefore, the model gives inflationary predictions identical with
those of the pure quadratic CI, although with \sub\ inflaton
values; thus, it is clearly in tension with the recent data
\cite{kdust,gws2}.


\newpage


\begin{thebibliography}{10} {\baselineskip 2.pt \ftn

\bibitem{chaotic} A.D. Linde, \plb{129}{1983}{177}.

\bibitem{wmap} G. Hinshaw \etal\ [WMAP Collaboration], {\sl Astrophys. J. Suppl. }{\bf 208} 19 (2013)  [\arxiv{1212.5226}].

\bibitem{plin} P.A.R.~Ade {\it et al.}  [\plk\ Collaboration], {\sl Astron.\ Astrophys.}\  {\bf 571}, A16 (2014)
[\arxiv{1303.5076}];\\ \arxiv{1303.5082}; {\tt
http://www.esa.int/Planck}.




%\bibitem{sb} W.~Buchm\"uller, E.~Dudas, L.~Heurtier and C.~Wieck, \jhep{09}{2014}{053} [\arxiv{1407.0253}].
  %%CITATION = ARXIV:1407.0253;%%



\bibitem{gws} P.A.R. Ade \etal\ [\bicep\ Collaboration], {\sl Phys. Rev. Lett. } {\bf 112}, 241101 2014) [\arxiv{1403.3985}].


\bibitem{oss1} N.~Okada, V.N. \c{S}eno\u{g}uz and Q.~Shafi, \arxiv{1403.6403}.

\bibitem{oss} T.~Kobayashi and O.~Seto,  {\sl Phys.\ Rev.\ D} {\bf 89}, 103524 (2014)
[\arxiv{1403.5055}]; \\ S.M.~Boucenna, S.~Morisi, Q.~Shafi and
J.W.F.~Valle, \prd{90}{2014}{055023} [\arxiv{1404.3198}].


\bibitem{shift1}  M.~Kawasaki, M.~Yamaguchi and T.~Yanagida,
\prl{85}{2000}{3572} [\hepph{0004243}];\\  P.~Brax and J.~Martin,
{\sl Phys.\ Rev.\ D\ }{\bf 72} 023518 (2005) [\hepth{0504168}];
\\S. Antusch, K. Dutta, P.M. Kostka, \plb{677}{2009}{221}
[\arxiv{0902.2934}];\\ R.~Kallosh, A.~Linde and T.~Rube,
\prd{83}{2011}{043507} [\arxiv{1011.5945}];\\   T.~Li, Z.~Li and
D.V.~Nanopoulos, \jcap{02}{2014}{028} [\arxiv{1311.6770}];\\
K.~Harigaya and T.T.~Yanagida, \plb{734}{2014}{13}
[\arxiv{1403.4729}]; \\ R.~Kallosh, A.~Linde and A.~Westphal,
\prd{90}{2014}{023534} [\arxiv{1405.0270}];\\ A.~Mazumdar,
T.~Noumi and M.~Yamaguchi, \prd{90}{2014}{043519}
[\arxiv{1405.3959}];\\  C.~Pallis and Q.~Shafi,
\plb{736}{2014}{261} [\arxiv{1405.7645}].
  %%CITATION = ARXIV:1405.7645
  %%CITATION = ARXIV:1405.0270;%%
%%CITATION = ARXIV:1011.5945;%%
%%CITATION = ARXIV:1403.4729;%%
%%CITATION = ARXIV:1311.6770;%%


\bibitem{rStar} S.~Ferrara, A.~Kehagias and A.~Riotto,
{\sl Fortsch.\ Phys.\  }{\bf 62}, 573 (2014) [\arxiv{1403.5531}];
\\ R.~Kallosh \etal\ \jcap{07}{2014}{053}
[\arxiv{1403.7189}]; \\ K.~Hamaguchi, T.~Moroi and
T.~Terada, \plb{733}{2014}{305} [\arxiv{1403.7521}]; \\
S.~Ferrara, A.~Kehagias and A.~Riotto, \arxiv{1405.2353}.
%A.~Linde, B.~Vercnocke and W.~Chemissany


\bibitem{rEllis} J.~Ellis, M.~Garcia, D.~Nanopoulos and
K.~Olive, \jcap{05}{2014}{037} [\arxiv{1403.7518}]; \\
J.~Ellis, M.~Garcia, D.~Nanopoulos and K.~Olive,
\jcap{08}{2014}{044} [\arxiv{1405.0271}].
  %%CITATION = ARXIV:1405.0271;%%



\bibitem{gws1} H. Liu, P. Mertsch and S. Sarkar, {\sl Astrophys. J. } {\bf 789}, L29 (2014)
[\arxiv{1404.1899}]; \\ J.~Martin, C.~Ringeval, R.~Trotta and
V.~Vennin, \prd{90}{2014}{063501} [\arxiv{1405.7272}];\\ R.
Flauger, J.C. Hill and D.N. Spergel, \jcap{08}{2014}{039}
[\arxiv{1405.7351}]; \\ M.~Cort\^es, A.R.~Liddle and D.~Parkinson,
\arxiv{1409.6530}.
  %%CITATION = ARXIV:1405.7272;%%


\bibitem{gws2} M.J.~Mortonson and U.~Seljak, \jcap{10}{2014}{035} [\arxiv{1405.5857}].
  %%CITATION = ARXIV:1405.5857;%%



\bibitem{pdust} R. Adam \etal\ [\plk\ Collaboration],
\arxiv{1409.5738}.



\bibitem{kdust} C. Cheng, Q. G. Huang and S. Wang,
\jcap{12}{2014}{044} [\arxiv{1409.7025}]; \\ L.~Xu,
\arxiv{1409.7870}.

\bibitem{circ} V.N. \c{S}eno\u{g}uz and Q.~Shafi, \plb{668}{2008}{6}
[\arxiv{0806.2798}].
%%CITATION = PHLTA,B668,6;%%

\bibitem{Qenq} K.~Enqvist and M.~Kar\v ciauskas, \jcap{02}{2014}{034} [\arxiv{1312.5944}].


\bibitem{nMCI} C. Pallis and Q. Shafi, \prd{86}{2012}{023523} [\arxiv{1204.0252}].
%%CITATION = ARXIV:1204.0252;%%

\bibitem{nmi} C. Pallis, \plb{692}{2010}{287} [\arxiv{1002.4765}].
 %%CITATION = ARXIV:1002.4765;%%


\bibitem{roest} R. Kallosh, A. Linde and D. Roest, {\sl Phys. Rev. Lett. }{\bf 112}, 011303 (2014)
[\arxiv{1310.3950}].



\bibitem{talk} C. Pallis, {\sl PoS Corfu \textbf{2012}}, 061  (2013)
[\arxiv{1307.7815}].
%%CITATION = ARXIV:1307.7815;%%

\bibitem{shap} F.L.~Bezrukov, A.~Magnin and M.~Shaposhnikov, \plb{675}{2009}{88}
[\arxiv{0812.4950}]; \\ D.P.~George, S.~Mooij and M.~Postma,
\jcap{02}{2014}{024} [\arxiv{1310.2157}]; \\  J.L.~Cook,
L.M.~Krauss, A.J.~Long and S.~Sabharwal, \prd{89}{2014}{103525}
[\arxiv{1403.4971}].
  %%CITATION = ARXIV:1310.2157;%%
 %``Standard Model Higgs boson mass from inflation,''
  %``Is Higgs Inflation Dead?,''


\bibitem{ld}  N.~Okada, M.U.~Rehman and Q.~Shafi, \prd{82}{2010}{043502} [\arxiv{1005.5161}]; \\
N.~Okada, M.U.~Rehman and Q.~Shafi, \plb{520}{2011}{701}
[\arxiv{1102.4747}].
%%CITATION = ARXIV:1007.1672;%%
  %%CITATION = ARXIV:1005.5161;%%


%\bibitem{prism}
  %``PRISM (Polarized Radiation Imaging and Spectroscopy Mission): An Extended White Paper,''
  %``PRISM (Polarized Radiation Imaging and Spectroscopy Mission): A White Paper on the Ultimate Polarimetric Spectro-Imaging of the Microwave and Far-Infrared Sky,''
%P.~Andre {\it et al.}  [PRISM Collaboration],
%\arxiv{1306.2259};\\

\bibitem{core} F.R.~Bouchet {\it et al.}  [COrE Collaboration],
\arxiv{1102.2181}; {\ftn\tt http://www.core-mission.org} \\
P.~Andre {\it et al.}  [PRISM Collaboration], \jcap{02}{2014}{006}
[\arxiv{1310.1554}].

  %``COrE (Cosmic Origins Explorer) A White Paper,''

\bibitem{pixie} A.~Kogut {\it et al.}, \jcap{07}{2011}{025} [\arxiv{1105.2044}].
  %``The Primordial Inflation Explorer (PIXIE): A Nulling Polarimeter for Cosmic Microwave Background Observations,''

\bibitem{lite} T.~Matsumura {\it et al.},  {\sl Journal of Low Temperature Physics} {\bf
176}, 733 (2014) [\arxiv{1311.2847}].
  %``Mission design of LiteBIRD,''
  %%CITATION = ARXIV:1311.2847;%%

\bibitem{nIG}  C.~Pallis, \jcap{08}{2014}{057}
[\arxiv{1403.5486}]; \\ C.~Pallis, \jcap{10}{2014}{058}
[\arxiv{1407.8522}].
  %%CITATION = ARXIV:1403.5486;%%
  %%CITATION = ARXIV:1407.8522;%%

\bibitem{plnext} J. Tauber \etal\ [\plk\ Collaboration],
\astroph{0604069}.

\bibitem{cmbpol} D.~Baumann {\it et al.}  [CMBPol Study Team Collaboration], {\sl AIP Conf.\ Proc.}\  {\bf 1141}, 10 (2009)
[\arxiv{0811.3919}].
  %``CMBPol Mission Concept Study: Probing Inflation with CMB Polarization,''


\bibitem{cutoff} J.L.F.~Barbon and J.R.~Espinosa,
\prd{79}{2009}{081302} [\arxiv{0903.0355}];\\
C.P.~Burgess, H.M.~Lee and M.~Trott, \jhep{07}{2010}{007}
[\arxiv{1002.2730}];\\ M.P.~Hertzberg, \jhep{11}{2010}{023}
[\arxiv{1002.2995}].

\bibitem{cutof} F.~Bezrukov, A.~Magnin, M.~Shaposhnikov and
S.~Sibiryakov,\\ \jhep{016}{2011}{01} [\arxiv{1008.5157}].
  %%CITATION = ARXIV:1310.7410;%%

\bibitem{riotto} A.~Kehagias, A.M.~Dizgah and A.~Riotto,
\prd{89}{2014}{043527} [\arxiv{1312.1155}].
  %``Comments on the Starobinsky Model of Inflation and its Descendants,''


\bibitem{Maeda} Y. Fujii and K. Maeda,
\textit{The Scalar-Tensor Theory of Gravitation} (Cambridge,
2003).

\bibitem{linde1} M.B.~Einhorn and D.R.T.~Jones,
\jhep{03}{2010}{026} [\arxiv{0912.2718}]; \\ H.M.~Lee,
\jcap{08}{2010}{003} [\arxiv{1005.2735}]; \\ S.~Ferrara \etal,
\prd{83}{2011}{025008} [\arxiv{1008.2942}];\\ C.~Pallis and
N.~Toumbas, \jcap{02}{2011}{019} [\arxiv{1101.0325}].
%%CITATION = ARXIV:1101.0325;%%

\bibitem{ketov} S.V.~Ketov and T.~Terada, \arxiv{1408.6524}; S.~Aoki and Y.~Yamada, \arxiv{1409.4183}.

\bibitem{nil} I.~Antoniadis, E.~Dudas, S.~Ferrara and A.~Sagnotti, {\sl Phys.\ Lett.\ B } {\bf 733}, 32 (2014)
[\arxiv{1403.3269}]; \\ S.~Ferrara, R.~Kallosh and A.~Linde,
\arxiv{1408.4096};\\ R.~Kallosh and A.~Linde, \arxiv{1408.5950};\\
G.~Dall'Agata and F.~Zwirner, \arxiv{1411.2605}.
  %``On sgoldstino-less supergravity models of inflation,''
  %%CITATION = ARXIV:1408.5950;%%
  %%CITATION = ARXIV:1408.4096

\bibitem{eno7} J.~Ellis, D.V.~Nanopoulos and K.A.~Olive,
  %``Starobinsky-like Inflationary Models as Avatars of No-Scale Supergravity,''
   \jcap{10}{2013}{009}  [\arxiv{1307.3537}].
  %%CITATION = ARXIV:1307.3537;%%

\bibitem{noscale} E.~Cremmer, S.~Ferrara, C.~Kounnas and D.V.~Nanopoulos,
{\sl Phys.\ Lett.\ B } {\bf 133}, 61 (1983); \\ J.R.~Ellis,
A.B.~Lahanas, D.V.~Nanopoulos and K.~Tamvakis, {\sl Phys.\ Lett.\
B } {\bf 134}, 429 (1984).


\bibitem{review} D.H.~Lyth and
A.~Riotto, {\sl Phys.\ Rept.} {\bf 314}, 1 (1999) [{\tt\ssz
hep-ph/9807278}];  \\ G. Lazarides, {\sl J. Phys. Conf. Ser.} {\bf
53}, 528 (2006) [{\tt\ssz hep-ph/0607032}]; \\ A.~Mazumdar and
J.~Rocher, {\sl Phys.\ Rept. }{\bf 497}, 85 (2011) [\arxiv{1001.0993}];  \\
J.~Martin, C.~Ringeval and V.~Vennin, {\sl Physics of the Dark
Universe} {\bf 5-6}, 75 (2014)  [\arxiv{1303.3787}].
%%CITATION = HEP-PH 0607032;%%



%A.R. Liddle and S.M. Leach, \prd{68}{2003}{103503}
%[\astroph{0305263}]; \\

\bibitem{hinova} C. Pallis, {\sl ``High Energy
Physics Research Advances''}, edited by T.P. Harrison and R.N.
Gonzales (Nova Science Publishers Inc., New York, 2008)
[\arxiv{0710.3074}].
%%CITATION = ARXIV:0710.3074;%%


\bibitem{liddle} L.~Dai, M.~Kamionkowski and J.~Wang,
\prl{113}{2014}{041302} [\arxiv{1404.6704}]; \\  J.B.~Munoz and
M.~Kamionkowski, \arxiv{1412.0656}.
%%CITATION = ARXIV:1404.6704;%%

\bibitem{cw} S.R.~Coleman and E.J.~Weinberg, \prd{7}{1973}{1888}.




\bibitem{reheating} Y.~Watanabe and E.~Komatsu, \prd{75}{2007}{061301}
[\grqc{0612120}];\\ Y.~Watanabe, \prd{83}{2011}{043511}
[\arxiv{1011.3348}].




\bibitem{quin} C. Pallis, \npb{751}{2006}{129} [\hepph{0510234}].
%%CITATION = NUPHA,B751,129;%%


\bibitem{aroest}  R.~Kallosh, A.~Linde and D.~Roest,
\jhep{11}{2013}{198} [\arxiv{1311.0472}]; \\  R.~Kallosh, A.~Linde
and D.~Roest, \jhep{08}{2014}{052} [\arxiv{1405.3646}].





%\bibitem{linde} R.~Kallosh and A.~Linde,
  %``Superconformal generalizations of the Starobinsky model,''
%  \jcap{06}{2013}{028} [\arxiv{1306.3214}].




\bibitem{r2} C.~Pallis, \jcap{04}{2014}{024} [\arxiv{1312.3623}].
  %%CITATION = ARXIV:1312.3623;%%


%\bibitem{cwSUGRA} S. Ferrara, C. Kounnas and F. Zwirner,
%\npb{429}{1994}{589}; \\ Erratum-\ibid{B433}{1995}{255}
%[\hepth{9405188}].


\bibitem{nmH} C.~Pallis and N.~Toumbas, \jcap{12}{2011}{002}
[\arxiv{1108.1771}].
  %%CITATION = ARXIV:1108.1771;%%

\bibitem{rRiotto} A.~Kehagias and A.~Riotto, \prd{89}{2014}{101301}
[\arxiv{1403.4811}].


\bibitem{rlimit} S.~Antusch and D.~Nolde, \jcap{05}{2014}{035} [\arxiv{1404.1821}].
  %%CITATION = JCAPA,0608,004;%%
 %%CITATION = PRLTA,78,1861;%%



\bibitem{rfhi} M.~Civiletti, C.~Pallis and Q.~Shafi,
{\sl Phys.\ Lett.\ B } {\bf 733}, 276 (2014) [\arxiv{1402.6254}].
%%CITATION = ARXIV:1404.1821;%%


\bibitem{lyth} D.H.~Lyth, \prl{78}{1997}{1861}
[\hepph{9606387}];\\ R.~Easther, W.H.~Kinney and B.A.~Powell,
\jcap{08}{2006}{004} [\astroph{0601276}].
  %%CITATION = JCAPA,0608,004;%%
 %%CITATION = PRLTA,78,1861;%%


}

\end{thebibliography}
\end{document}